\documentclass{jfm}
\usepackage{graphicx}
\usepackage{todonotes}
\usepackage{epstopdf, epsfig}
\usepackage{amsmath}
\usepackage{xcolor}
\usepackage{lipsum}  
\usepackage{hyperref}
\usepackage[hang,splitrule,symbol]{footmisc}

\usepackage{multirow}
\usepackage{lineno}
\usepackage{mathrsfs}

\usepackage{float}

\usepackage{xspace}

\makeatletter
\DeclareRobustCommand\onedot{\futurelet\@let@token\@onedot}
\def\@onedot{\ifx\@let@token.\else.\null\fi\xspace}

\def\eg{\emph{e.g}\onedot} 
\def\ie{\emph{i.e}\onedot}

\def\etal{\emph{et al}\onedot}
\makeatother

\shorttitle{Wake meandering for a FOWT under side-to-side motion}
\shortauthor{Zhaobin Li, Guodan Dong and Xiaolei Yang}

\title{Onset of wake meandering for a floating offshore wind turbine under side-to-side motion}

\author{Zhaobin Li\aff{1,2},  Guodan Dong\aff{1,2} 
 \and Xiaolei Yang\aff{1,2}\corresp{\email{xyang@imech.ac.cn}}}

\affiliation{\aff{1}The State Key Laboratory of Nonlinear Mechanics, Institute of Mechanics, Chinese Academy of Sciences, Beijing 100190, China
\aff{2}School of Engineering Sciences, University of Chinese Academy of Sciences, Beijing 100049, China}

\begin{document}

\maketitle

\begin{abstract}

Wind turbine's wake, being convectively unstable, may behave as an amplifier of upstream perturbations and make the downstream turbine experience strong inflow fluctuations. In this work, we investigate the effects of the side-to-side motion of a floating offshore wind turbine (FOWT) on wake dynamics using large-eddy simulation and linear stability analysis (LSA) on the NREL 5MW baseline offshore wind turbine. Simulation results reveal that the turbine motion can lead to wake meandering for motion frequencies with the Strouhal number $St = fD/U_\infty \in (0.2,0.6)$ (where $f$ is the motion frequency, $D$ is the rotor diameter, and $U_\infty$ is the incoming wind speed), which lie in the range of the natural roll frequencies of common FOWT designs. This complements the existing wake meandering mechanism, that the side-to-side motion of a FOWT can be a novel origin for the onset of wake meandering. The amplitude of the induced wake meandering can be one order of magnitude higher than the initial perturbation for the most unstable frequencies. The probability density function of the spanwise location of the instantaneous wake centers is observed having two peaks on the time-averaged wake boundaries and a trough near the time-averaged wake centerline, respectively. It is also found that the LSA can predict the least stable frequencies and the amplification factor with acceptable accuracy for motion amplitude $0.01D$. Effects of non-linearity are observed when motion amplitude increases to $0.04D$, for which the most unstable turbine oscillations shift slightly to lower frequencies and the amplification factor decreases.

\end{abstract}

\maketitle

\section{Introduction}

When a wind turbine generates electricity from wind, it leaves behind a wake featured by low wind speed and high turbulence intensity. In a wind farm, the wake can result in a decrease of energy production and an increase of fatigue loads for downstream turbines. For this reason, to understand and to control the wake for mitigating these negative impacts have become one of the major challenges faced by the wind energy industry and the wind energy science community  \citep{van_wingerden_expert_2020,veers2019grand,meneveau2019big,porte2020wind}. For floating offshore wind turbines (FOWTs), the motion of the platform further complicates the dynamics of wind turbine wakes. In this study, we investigate the effects of the side-to-side motion of a FOWT on the dynamics of wind turbine wakes, and show that such motion of a FOWT can trigger wake meandering and help the recovery of wind turbine wakes. 
%

In the literature, a number of studies
have been conducted to identify the features of wind turbine wakes and to reveal the underlying physics using field measurements  \citep{mechali2006wake,hong2014natural,sun2020review}, wind tunnel experiments  \citep{medici2006measurements,chamorro2009wind,espana2012wind,bastankhah_experimental_2016}, and numerical simulations  \citep{kang2014onset,yang2016coherent,yang2018large,li2020evaluation}. 
An essential feature discovered by these work is the unsteady nature of the wind turbine wake, especially its low-frequency, large-scale lateral motion, referred to as wake meandering  \citep{medici2006measurements,yang2018large,li2020evaluation,larsen2007Meandering}. From the time-averaged point of view, the wake meandering spreads the wake in a large region and reduces the time-averaged velocity deficit  \citep{li2020similarity}. On the other hand, the wake meandering results in unsteady loads on downstream turbines since it frequently alters the incoming wind of the downwind turbines between full wake, partial wake and the freestream  \citep{larsen2008wake}. Two mechanisms have been proposed to explain the origin of wake meandering: i) the large-eddy mechanism, and ii) the shear layer instability mechanism. The first mechanism regards the wind turbine wakes as passive scalars that are transported by eddies larger than the rotor diameter in the atmospheric boundary layer (ABL) so that the wake trajectory is deformed as travelling downstream. With Taylor frozen flow hypothesis, the first mechanism is employed for developing the dynamic wake meandering model (DWM)  \citep{larsen2008wake,madsen2010calibration,keck2014atmospheric}. On the other hand, the second mechanism considers the wake meandering as the intrinsic property of the wake flow, which is triggered by the selective amplification of upstream disturbance. Recently,  computational simulations of  \citet{yang2018large} and field measurements of \citet{heisel2018spectral} both confirmed the coexistence of both mechanism at utility scale, and shows that the large-eddies mechanism mainly contributes to very low frequent wake meandering $\textrm{St} \ll 0.1$ while the shear layer mechanism contributes to the wake meandering in the range of $0.1< \textrm{St} <0.3$, respectively.

Examination on the two meandering mechanisms has been carried out using field measurements, wind tunnel experiments and numerical simulations. To verify the conjecture of the large-eddy mechanism, \citet{trujillo2011light} compared the prediction of the DWM with lidar measurements of instantaneous wake center positions behind a 95 kW wind turbine and found a cross-covariance about 0.58. Moreover,  \citet{espana2011spatial,espana2012wind} studied the wake meandering process by modeling the wind turbine as a porous disk in a wind tunnel and confirmed that the wake meandering is dominated by eddies with integral length larger than the rotor diameter. The shear layer instability mechanism is often examined by comparing the measured/computed wake meandering frequency with the typical frequency of bluff bodies. \citet{medici2006measurements} carried out wind tunnel experiments of a miniature wind turbine (rotor diameter $D=0.18~$m), and observed low-frequent periodic wake fluctuations with $ 0.12 < \textrm{St} < 0.25$ (varying with rotor thrust and yaw angle). Later, \citet{chamorro2010effects} found different thermal stability conditions can alter the meandering frequency in the range of $0.33<St<0.40$ for a miniature wind turbine. \citet{barlas2016roughness} revealed the effect of inflow turbulence on the wake meandering and showed that the wake meandering happens at $\textrm{St} \approx 0.25 $ for ambient turbulence intensity TI = 6.8\% but is suppressed when TI is increased to 14\%. 
\citet{iungo2013linear} performed linear stability analysis (LSA) on the time-averaged flow from a wind tunnel experiment, and found that the measured wake meandering frequencies match well the most amplified frequency computed from LSA. \citet{mao2018far} scrutinised the meandering of the wake behind a porous disk triggered by atmospheric eddies and found the most amplified perturbation falls in the frequency range of $\textrm{St} \in \left[0.25, 0.63 \right]$. \citet{gupta2019low} investigated the far wake meandering behind a tidal turbine with inflow velocity magnitude periodically varying using large eddy simulations, and developed a low-order model based on the complex Ginzburg–Landau (CGL) equation.

For FOWTs, we speculate that the disturbances introduced by the platform motion may trigger wake meandering. 
In the literature, distortion of near wake structures by the surge and pitch motions of the platform was observed in computational results obtained using the free vortex method  \citep{farrugia2016study,lee2019effects} and the unsteady Reynolds-averaged Navier-Stokes (URANS) method in recent studies  \citep{tran2016cfd,liu2016investigation}. Such influence of platform pitch and surge motions on the wake was also observed in several wind tunnel experiments but with controversial results  \citep{rockel2014experimental,rockel2017dynamic,schliffke2020wind, fu2019wake,fu2020phase}. For instance, \citet{rockel2014experimental, rockel2017dynamic} considered the effects of platform pitch motion on the far wake evolution and observed a decrease of wake replenishing due to a time-averaged upward wake deflection. \citet{schliffke2020wind} reported no apparent influence on the wake at $4.6$ rotor-diameter downstream even with surge motions of large amplitude according to their wind tunnel measurement of an actuator disk type wind turbine model. As seen, most work have been focused on the surge and the pitch (\ie{},  fore-aft motions) of the platform. However, waves may also induce side-to-side platform motions since they are not always aligned with the wind direction  \citep{wanninkhof1992relationship},  
and such oblique waves are able to induce side-to-side turbine motions as large as the fore-aft motions  \citep{lyu2019effects}.  Recently, \citet{fu2019wake} measured the wake of a miniature wind turbine rolling at relatively low Strouhal numbers $\textrm{St} < 0.02$ in a wind tunnel. However, they found that effect of roll motion only resides in the near wake and become negligible at $7$ rotor-diameter downstream, different from wake meandering. To the best of the authors' knowledge, no systematic studies have been carried out to examine the possibility of wake meandering induced by the motion of a FOWT. 

As a first step, this work focuses on the side-to-side motion of the platform considering that it is in the lateral direction of the wake and may have the largest potential to trigger wake meandering. The NREL 5MW baseline offshore wind turbine parametrized with the actuator surface model is simulated using large-eddy simulation. The side-to-side motion of the platform is simplified as imposed harmonic motion. A wide range of amplitudes and frequencies are selected in the LES together with local LSA to look for those that can induce wake instability. 

The rest of the paper is organized as follows. The numerical methods and simulation setups are presented in Section 2. Then simulation results are shown in Section 3, including the wake statistics and the local LSA.  At last the discussion and conclusion are given in Sections 4 and 5, respectively. 

\section{Numerical methods and simulation setups}

\subsection{Flow solver}

The air flow around the wind turbine is simulated using the LES module of the Virtual Flow Simulator code, VFS-Wind  \citep{yang2015VFS}. The air is assumed to be an incompressible Newtonian fluid, which has a constant density and viscosity. The flow is governed by the filtered incompressible Navier-Stokes equations as follows, 
%
%
\begin{align}
	J \frac{ \partial U^j }{ \partial \xi^j } & =0, \label{eqn:cnt}\\
	\frac{1}{J}\frac{\partial U^i}{\partial t} & = \frac{\xi^i_l}{J}\left( -\frac{\partial }{\partial \xi^j} \left(U^ju_l\right)+\frac{\mu}{\rho}\frac{ \partial }{\partial \xi^j}\left(\frac{g^{jk}}{J}\frac{\partial u_l}{\partial \xi^k}\right) -\frac{1}{\rho}\frac{\partial}{\partial \xi^j}\left( \frac{\xi^j_l p}{J}\right)-\frac{1}{\rho}\frac{\partial \tau_{lj}}{\partial \xi^j} +f_l\right), \label{eqn:ns}
\end{align}
where $i,j,k,l=\{1,2,3\}$ are tensor indices,  $\xi^i$ are the curvilinear coordinates which are related to the Cartesian coordinates $x_l$ with the transformation metrics $\xi^i_l  = \partial \xi^i/\partial x_l $.  $J$  denotes the Jacobian of the geometric transformation, $U^i = \left(\xi^i_l / J\right)u_l$ are the contravariant volume flux with $u_l$ the velocity in Cartesian coordinates. $\mu$ denotes the fluid dynamic viscosity and $\rho$ is the fluid density. $g^{jk} = \xi^j_l\xi^k_l$ are the components of the contravariant metric tensor. The pressure is denoted by $p$ and $f_l$ are body forces to account for the effects of wind turbines on the flow. $\tau_{ij}$ in the momentum equation is the subgrid-scale stress \citep{Smagorinsky},
\begin{equation}
	\tau_{ij}-\frac{1}{3}\tau_{kk}\delta_{ij}=-\mu_t \overline{S_{ij}},
\end{equation}
where $\overline{S_{ij}}$ is the filtered strain-rate tensor, $\overline{(\cdot)}$ denotes the grid filtering operation and $\mu_t$ is the eddy viscosity computed by 
	\begin{equation}
	\mu_t = C_s \Delta^2 |\overline{S}|,
	\end{equation}
where $\Delta$ is the filter width. $ |\overline{S}| = (2 \overline{S_{ij}} \overline{S_{ij}})^{1/2} $  is the magnitude of the strain-rate tensor and $C_s$ is the Smagorinsky constant computed via a dynamic procedure   \citep{germano1991subGridModel}.

The governing equations are discretized on a structured curvilinear grid. The second-order central differencing scheme is employed for the spatial discretization together with a second-order fractional step scheme  \citep{ge2007numerical} for the temporal integration. The momentum equation is solved with a matrix-free Newton-Krylov method  \citep{knoll2004jacobian}. The pressure Poisson equation derived from the continuity constraint, is solved using the Generalized Minimal Residual (GMRES) method with an algebraic multi-grid acceleration  \citep{saad1993flexible}.

\subsection{Wind turbine modeling}
A class of well-validated actuator surface model  \citep{yang2018ASMethod} is employed to model the rotor and the nacelle of the wind turbine. It represents a rotor blade with a simplified two dimensional surface defined by the chord length and the twist angle, with distributed forces computed from the lift and drag ($\mathbf{L}$ and $\mathbf{D}$) determined as follows:
\begin{equation}
\mathbf{L} = \frac{c}{2}\rho C_\text{L}|V_\text{ref}|^2 \mathbf{e_\text{L}}
\end{equation}
and 
\begin{equation}
\mathbf{D} = \frac{c}{2}\rho C_\text{D}|V_\text{ref}|^2 \mathbf{e_\text{D}}, 
\end{equation}
where $c$ denotes the local chord length, $C_\text{L}$ and $C_\text{D}$ are the lift and the drag coefficients defined in 2D airfoil tables as a function of Reynolds number and angle of attack, $\textbf{e}_\text{L}$ and $\textbf{e}_\text{D}$ are the unit directional vectors of the lift and drag. $V_\text{ref}$ is the local incoming velocity relative to the moving blades averaged over the local chord width. A 3D stall delay model  \citep{du19983Dstall} and a tip loss correction  \citep{shen2005tiploss2} are applied.  \textbf{L} and \textbf{D} are employed to calculate the body force in equation \eqref{eqn:ns} by uniformly distributing the resultant force along the chord width as follows,
\begin{equation}
\mathbf{f} = (\mathbf{L}+\mathbf{D})/c.  
\end{equation}

In the actuator surface model for nacelle, the normal component of the force is computed by satisfying the non-penetration boundary condition, and the tangential component is computed using a specified friction coefficient and incoming velocity, respectively. The transfer of quantities between the actuator surface and the background grid nodes is achieved via a smoothed discrete delta function  \citep{yang2009Kernel}.
\subsection{Simulation setup}
In the present cases, the NREL offshore 5MW baseline wind turbine  \citep{jonkman2009definition} is employed. It has a rotor with diameter $D = 126~$m and a cuboidal shape nacelle ($2.3~\text{m} \times 2.3~\text{m} \times 14.2~\text{m}$). The simulation represents the rated operation condition with the inflow wind speed of $U_\infty = 11.4~\text{m/s}$ and tip speed ratio of 7. The Reynolds number based on the inflow velocity and the rotor's diameter is  $ \textrm{Re} \approx 9.6\times 10^7$. The simulations are conducted with an idealized uniform inflow to reflect the low turbulence offshore environment  \citep{bodini2020offshore} with the wind veer and shear neglected so that the side-to-side rotor motion is the only source of disturbances for the wake.

\begin{figure}
    \centering
    \includegraphics[width=0.8\textwidth]{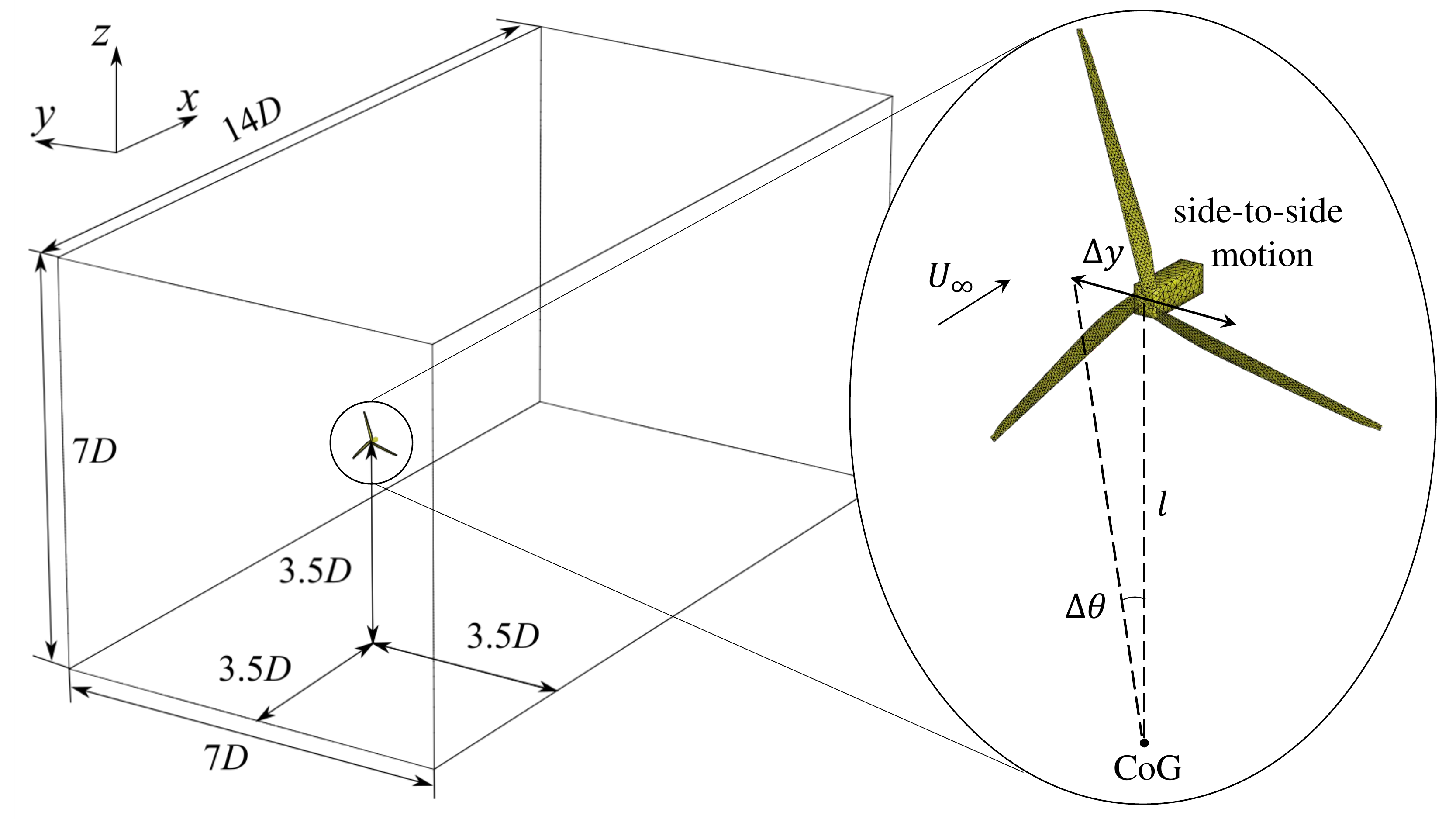}
    \caption{{
\color{black}Schematics of the computational domain and the rotor side-to-side motion.}}
    \label{fig:computationalDomain}
\end{figure}

The computational domain, showing in figure \ref{fig:computationalDomain}, is a cuboid with $L_x \times L_y \times L_z = 14D \times 7D \times 7D$, where $x,y \textrm{~and~} z$ denote the streamwise, horizontal and vertical directions, respectively. The turbine is placed $3.5D$ from the inlet and located at the center of the cross section. The origin of the coordinate system coincides with the wind turbine's hub center. The inflow velocity is imposed on the inlet boundary at $x = -3.5D$. Neumann condition for velocity ($\partial u_i/\partial x=0$) is applied on the outlet boundary at $x = 10.5D$. Free slip condition is imposed on the four lateral boundaries. The computational domain is discretized with a Cartesian grid. The streamwise grid is uniform with $\Delta x = D/20$. In $y$ and $z$ directions, the grid spacing is $ \Delta x = \Delta y = D/40 $ near the wake centerline, \ie{}, in the region $y \times z \in [-1.5D, 1.5D] \times [-1.5D, 1.5D]$. Out of this near
region, the grid is gradually stretched in both lateral directions. The total number of nodes is $N_x \times N_y \times N_z = 281 \times 141 \times 141 \approx 5.6$~M. Such a grid resolution is fine enough for the actuator model to predict the velocity deficit and turbulence intensity in turbine wakes with reasonable accuracy based on a previous grid refinement study  \citep{yang2015VFS}. Each simulation takes about 40 rotor revolutions to fully develop the wake initially. After this transitional stage, the instantaneous flow fields are saved for further investigation. 
\subsection{Side-to-side motions}
\begin{table}

\begin{tabular}{c@{\hskip 0.2in}c@{\hskip 0.2in}c@{\hskip 0.2in}c}
No. & Amplitude [$A$~(m)]   & Nondimensional Amplitude [$A/D$]  & Roll Amplitude [$\theta (^\circ)$] \\
1 & $1.26$     & 0.01   & 0.72 \\
2 & $2.52$    & 0.02    & 1.44 \\
3 & $5.04$    & 0.04    & 2.88 \\

\end{tabular}
\caption{Amplitudes of the simple harmonic motion of the rotor hub and the corresponding roll motion amplitudes. \label{tab:amplitudes} }

\end{table}

\begin{table}

\begin{tabular}{c@{\hskip 0.3in}c@{\hskip 0.35in}c@{\hskip 0.35in}c}
No.       & Strouhal Number [${St} = D/TU_\infty$]  & Period [$T$~(s)]   & Note\\
1        & 0.10      & 110.5           & Natural sway frequency\\
2         & 0.20      & 55.3          & $\downarrow$\\
3           & 0.25     & 44.2         & \\
4         & 0.30        & 36.8        & --------\\
5         & 0.35        & 31.6        & $\uparrow$ \\
6       & 0.40       & 27.6          & Natural roll frequency\\
7       & 0.45      & 24.6           & $\downarrow$\\
8          & 0.50       & 22.1        &  --------\\
9        & 0.60       & 18.4          & \\
10          & 0.80     & 13.8        &  $\uparrow$\\
11           & 1.00   & 11.1            & Ocean waves \\
\end{tabular}
\caption{{\color{black}The Strouhal numbers and the periods of the simple harmonic motions, the last column describes approximately the  corresponding ranges of the different physics.}   \label{tab:frequencies}}

\end{table}

In the simulations, the wind turbine is imposed with a simple harmonic motion in $y$ direction as shown in figure \ref{fig:computationalDomain}, which can either stem from the translation along the $y$ axis (sway motion) or the rotation along the $x$ axis (rolling motion) with respect to the center of gravity (CoG) of the floating wind turbine and the platform. For small roll motion, the side-to-side rotor hub displacement $\Delta y$ and the roll angle $\Delta \theta$ can be approximated with $\Delta y \approx l \Delta \theta$ where $l \approx 100~\textrm{m}$ is the distance from the hub to the CoG  \citep{robertson_2014_offshore}. In the present work, we only considered motions with relatively small amplitudes as listed in table \ref{tab:amplitudes} to ensure the hub acceleration are within the mechanical design requirement  \citep{nejad2019effect}. 

From the wave-structure interaction point of view, the motions of a FOWT can be divided into two groups, \ie{} those induced directly by waves and those at natural frequencies  \citep{jonkman_offshore_2010,robertson_2014_offshore}. For this reason, the present work selects a rather large frequency range to analyze the different influence of different motions from both groups, as shown in table \ref{tab:frequencies}. The high frequency part corresponds to the motions directly induced by the first-order wave forces, \ie{}, with periods approximately 10 $\sim$ 15 s at operational sea states \citep{toffoli2017types}. On the other hand, the natural frequencies, although change from case to case due to different turbine and platform designs, usually falls into a lower frequency range. For instance, the periods corresponding to the natural frequencies of roll and sway motions are 25$\sim$30 s and 100$\sim$120 s, respectively for currently available prototype floating platforms  \citep{jonkman_offshore_2010,robertson_2014_offshore}. These lower frequencies are also considered in the present work as shown in table~\ref{tab:frequencies}. 
\section{Results}

In this section, we first present the results from a specific case in section \ref{sect:result1}, for which the wake meandering is triggered by the side-to-side motion of the turbine. This case is employed to illustrate the difference between the wakes behind an oscillating turbine and a fixed turbine in detail. Moreover, this case is used to introduce the techniques employed for extracting the wake characteristics that will be further applied to all the other cases for investigating the influence of the motion amplitude and frequency in section \ref{sect:result2}. Section \ref{sect:result2} also presents the local LSA on the time-averaged wake behind a fixed wind turbine for predicting the wake meandering, with results compared with those obtained from LES cases. 

\subsection{Effects of wind turbine oscillations\label{sect:result1}}
In the case presented in this section, the frequency and amplitude of the turbine motion are $\textrm{St}=0.25$ and $A/D=0.04$, respectively. We will show that the meandering motion is triggered by the motion of the turbine for this case, and compare the wake of this oscillating turbine with that of a fixed turbine.  

\subsubsection{Instantaneous wake characteristics}
\begin{figure}
    \centering
    \includegraphics[width=\textwidth]{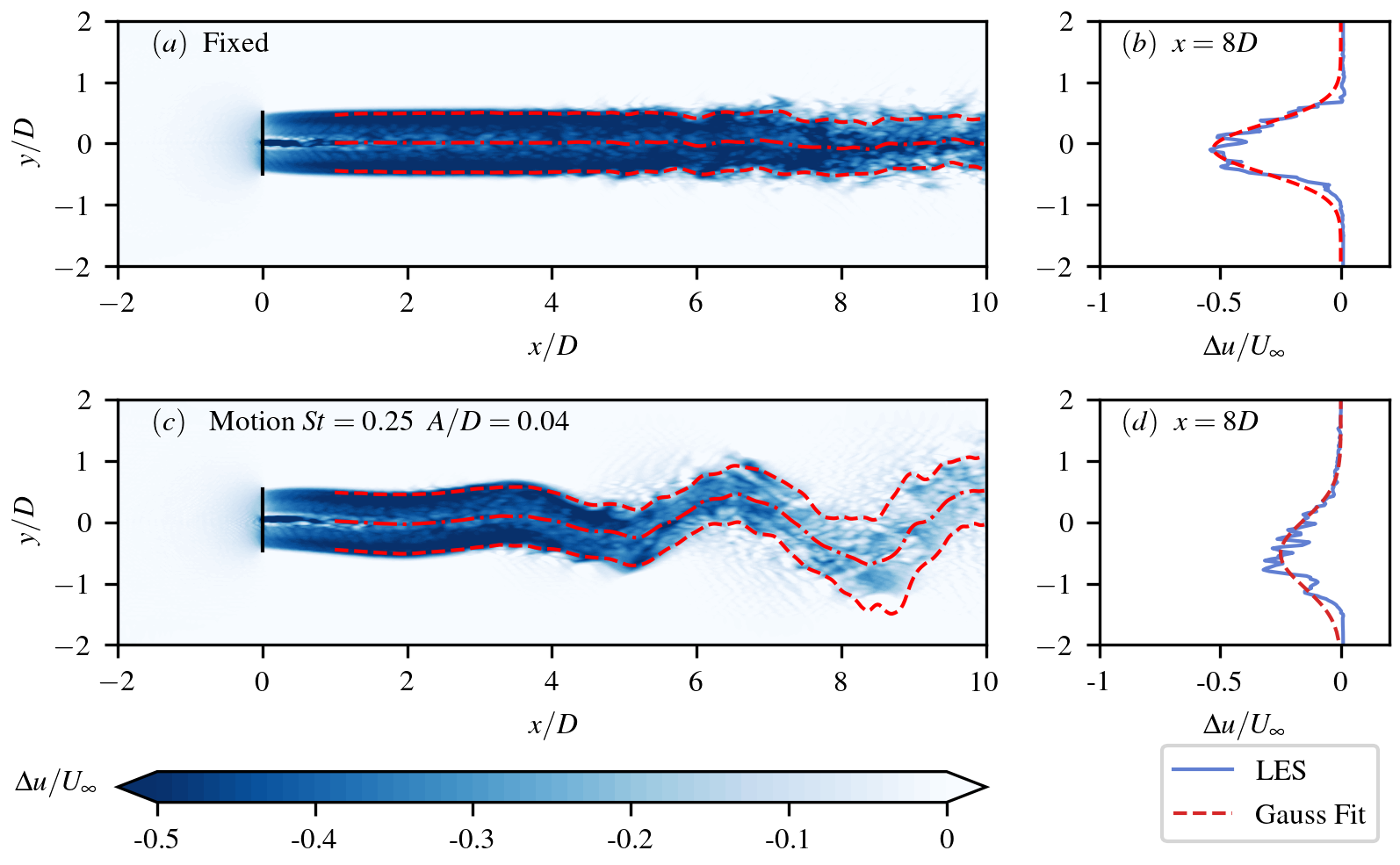}
    \caption{Instantaneous streamwise velocity deficit in the wake behind a fixed wind turbine (a) and a wind turbine subject to side-to-side simple harmonic motion of $\textrm{St} = 0.25$ and amplitude $A/D = 0.04$ (c). The wind turbines are illustrated with black solid lines. The red dash-dotted lines and the dashed lines indicate the center-lines and the wake width, respectively.  They are obtained with the Gaussian fit of the velocity deficit profile, as shown in (b) and (d) for the wakes of the fixed wind turbine and of the moving wind turbine, respectively.\label{fig:instantanousWakeCompare} }
\end{figure}
The influence of turbine side-to-side motion on the instantaneous wake is revealed in figure \ref{fig:instantanousWakeCompare}, where snapshots of the instantaneous wakes behind a fixed wind turbine and behind an oscillating one are compared. Figures \ref{fig:instantanousWakeCompare} (a) and (c) compare the  contours of the instantaneous streamwise velocity deficit with centerline and width of the wake extracted with a Gaussian fit defined as follows,   
\begin{equation}
     \Delta u(x,y) =  u(x,y) - U_\infty = \Delta u_c(x) \exp{\left(-\frac{(y-y_c(x))^2}{2 s^2(x)}\right)}, \label{eqn:selfsimilar}
\end{equation}
where $u(x,y)$ is the streamwise velocity and $s(x)$ is the standard deviation of the Gaussian distribution to be fitted with the velocity profile. The wake centerline position is defined as the peak location of the velocity deficit plotted with red dash-dotted lines. The wake width (red dashed lines) is defined as $r_{1/2}(x) = \sqrt{2\ln 2} s(x)$, which gives the distance from $y_c(x)$ to the position where $\Delta u = \displaystyle \frac{1}{2}\Delta u_c$. This fitting technique based on self-similarity, which is often applied to extract the  features of time-averaged wakes \citep{bastankhah_experimental_2016,xie2015self}, has been demonstrated being able to extract features of instantaneous wakes  \citep{li2020similarity}. 

As seen in figure \ref{fig:instantanousWakeCompare} (a), the centerline of the wake behind the fixed wind turbine is approximately straight in the near wake and fluctuates mildly in the far wake ($x>6D$). Moreover, the width of the wake behind the fixed turbine remains almost constant at all downstream locations, suggesting that the wake recovery is slow for the fixed turbine for the simulated case without inflow turbulence. In contrast, the oscillating wind turbine generates a wake with large-scale oscillations as shown in figure \ref{fig:instantanousWakeCompare} (c), revealing an significant impact of the side-to-side motion on turbine wakes. The meandering amplitude is found being approximately equal to the turbine diameter ($A \approx D$) in the far wake, while the amplitude of the side-to-side motion is only $A = 0.04D$. This rapid growth shows that the initial perturbation induced by the turbine motion is vastly amplified as traveling downstream. Besides the wake centerline oscillation, the velocity deficit in the far-wake is also significantly affected, being weaker behind the oscillating turbine when compared with the fixed turbine as shown in figures \ref{fig:instantanousWakeCompare} (a) and (c). This difference is more evident in figures \ref{fig:instantanousWakeCompare} (b) and (d), where the transverse profiles of the velocity deficit at downstream location $x=8D$ are plotted. The blue curves represent the instantaneous velocity deficit, and the red dashed lines are their Gaussian fits defined with equation \eqref{eqn:selfsimilar}. It is clear that the centerline velocity deficit $\Delta u_c$ is stronger and the wake width is narrower for the fixed wind turbine when compared with the oscillating wind turbine. As for the distribution of the velocity deficit, wake behind the fixed turbine is symmetric and can be fitted well with the Gaussian curve. The wake downstream of the oscillating wind turbine, on the other hand, shows obvious asymmetry. Despite the observed asymmetry, the essential features of the meandering wake, \ie{}, the centerline position, the velocity deficit, and the width, are well captured by the Gaussian fitting as shown in figures \ref{fig:instantanousWakeCompare} (c) and (d). 


\begin{figure}
    \centering
    \includegraphics[width=\textwidth]{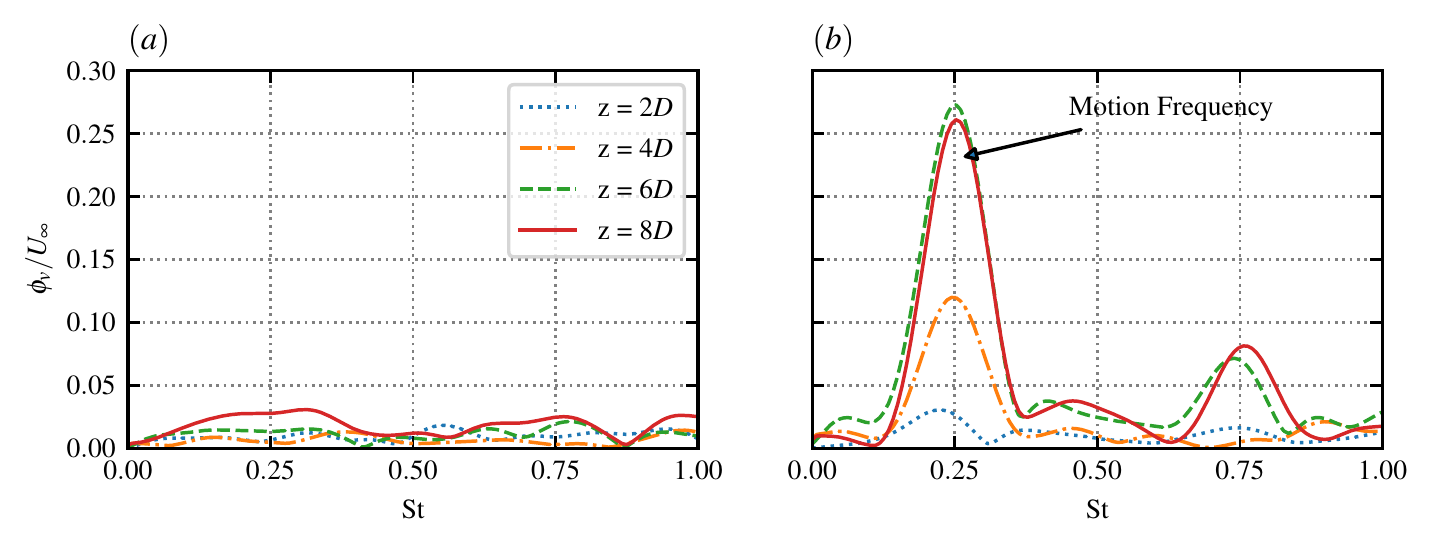}
    \caption{Frequency spectra of the spanwise velocity fluctuation in the wake behind the fixed wind turbine $(a)$ and the wind turbine subject to side-to-side FOWT motion with $\textrm{St} = 0.25,~A/D=0.04~ (b)$. Data are collected at four downstream locations ($x = \{2D,~4D,~6D,~8D\}$) along $y=0$. }
    \label{fig:velocityFrequencySpectra}
\end{figure}

The wakes downstream of the fixed and the oscillating wind turbines are further compared by examining the frequency spectra of the spanwise velocity fluctuation ($\phi_v$) in figure \ref{fig:velocityFrequencySpectra}. The velocity spectra ($\phi_v$) is computed with instantaneous spanwise velocities ($v$) recorded along the time-averaged wake centerline ($y=0$) at four turbine downstream locations ($x = \{2D,~4D,~6D,~8D\}$) using the discrete Fourier transform. As shown in figure \ref{fig:velocityFrequencySpectra} ($a$), the velocity behind the fixed wind turbine contains only small-scale fluctuations, which are influenced by the hub vortex the wake of the nacelle, and distributed over a wide range of frequencies. No obvious growth can be found for the fluctuation amplitude as traveling downstream. In contrast, the velocity spectra behind the oscillating wind turbine consist of a distinct peak at the frequency of the imposed motion ($\textrm{St}=0.25$). Moreover, the spectral level at $\textrm{St}=0.25$ grows significantly with downstream distances, revealing the onset of the instability of the wake, which significantly amplifies the disturbances initiated by the side-to-side motion, with the spectral level at other frequencies remaining approximately the same except for the trice of the imposed frequency. It is observed that the spectral level at the imposed frequency ceases to grow from $x=6D$ to $x=8D$, while the peak at the higher frequency ($\textrm{St} \approx 0.75$) start to appear at $x=6D$, which indicates the saturation of the growth of the imposed motion and the energy transfer from low to high frequencies via nonlinear effects  \citep{schmid2002stability,drazin2004hydrodynamic}. 

\begin{figure}
    \centering
    \includegraphics[width=\textwidth]{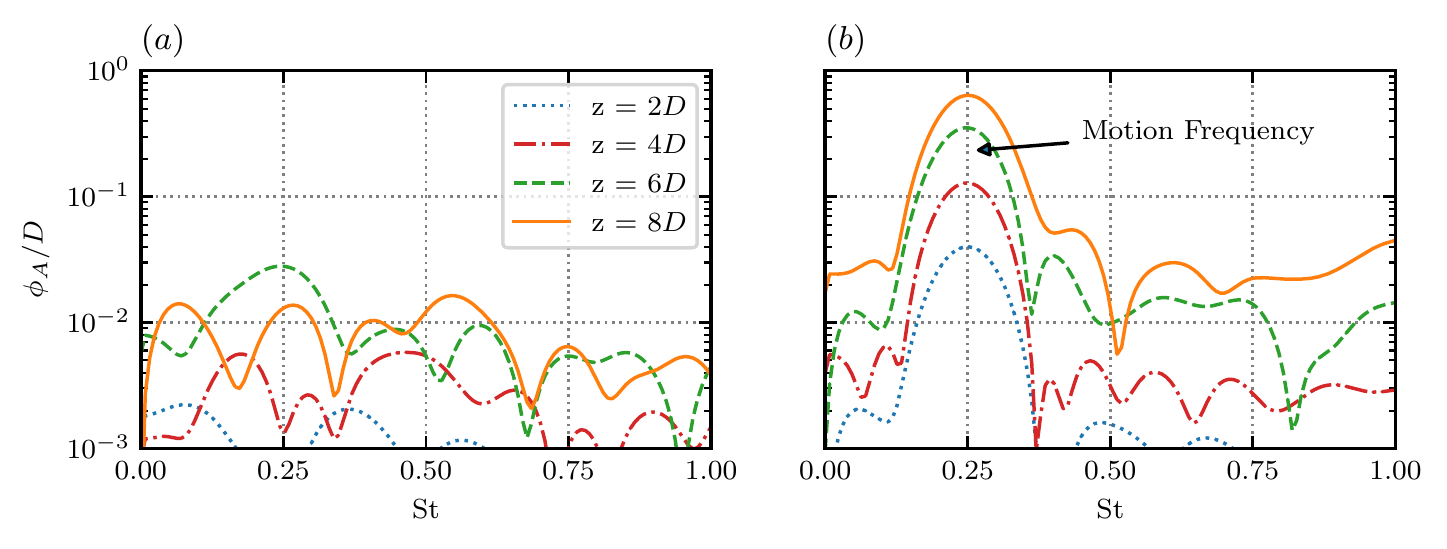}
    \caption{Frequency spectra of the wake center motion behind the fixed wind turbine $(a)$ and the wind turbine subject to side-to-side FOWT motion at $\textrm{St} = 0.25,~A/D=0.04~ (b)$. Data are collected at four downstream locations ($x = \{2D,~4D,~6D,~8D\}$) along $y=0$. }
    \label{fig:wakeCenterPositionSpectra}
\end{figure}

Similar to above analyses, the spectra of wake center displacement is computed and shown in figure \ref{fig:wakeCenterPositionSpectra}. It is observed that the amplitudes of wake center oscillations behind the oscillating wind turbine are one to two orders of magnitude higher than the fixed wind turbine at the frequency of the imposed side-to-side motion. For other frequencies, the spectral levels from the oscillating wind turbine are comparable to the fixed wind turbine, which does not have any distinct peak frequencies.
Furthermore, it is observed that the spectral level at the peak frequency grows monotonically as traveling downstream from the turbine and is comparable to the rotor diameter in the far wake ($x=6D, 8D$). 

\begin{figure}
    \centering
    \includegraphics[width=\textwidth]{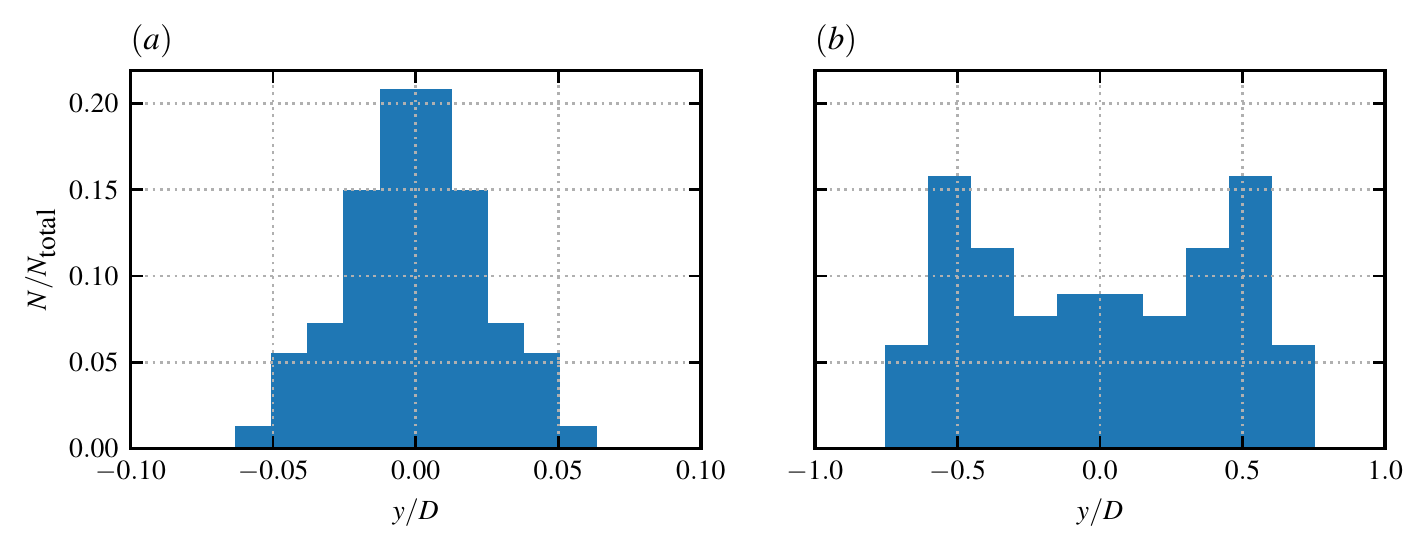}
    \caption{{\color{black}The probability density function of wake center positions at turbine downstream location $x=8D$ for the fixed wind turbine $(a)$ and the oscillating wind turbine $(b)$.}}
    \label{fig:wakeCenterlinePositionPDF}
\end{figure}

{\color{black}
Another interesting observation is that the spanwise distribution of instantaneous wake center positions becomes very different for the oscillating turbine when compared with the fixed one. These instantaneous wake centers are sampled from instantaneous flow fields at evenly distributed time steps. In figure \ref{fig:wakeCenterlinePositionPDF}, the PDF of instantaneous wake positions at turbine downstream location $x=8D$ are compared. As seen, the PDF for the wake centers of the fixed wind turbine as shown in figure \ref{fig:wakeCenterlinePositionPDF} (a) approximately follows the normal distribution located in a relatively narrow band with $y/D \in [-0.05,0.05] $ with a single peak located near $y=0$. In contrast, the wake centers behind the oscillating wind turbine spread in a much wider extend with $y \in (-0.7D, 0.7D)$, being more than 10 times larger than the fixed wind turbine. More importantly, the PDF shows a distinct shape with two peaks near $y=\pm 0.5D$ and a trough near $y=0$, which is significantly different from that observed for the fixed wind turbine  \citep{li2020similarity} and reflects the fact that the wake center spends more time at $y = \pm 0.5D$ than at $y=0$. It is noticed that this PDF is close to that of a simple harmonic motion with a trough in the equilibrium position and two peaks at extremities  \citep{tennekes1972firstCourse}, indicating the significant effect the turbine motion has on the frequency of the meandering motion. From the practical point of view, this PDF implies that the wake of an oscillating wind turbine may partially shelter the downstream turbine more often than a fixed wind turbine. }
\begin{figure}
    \centering
    \includegraphics[width=\textwidth]{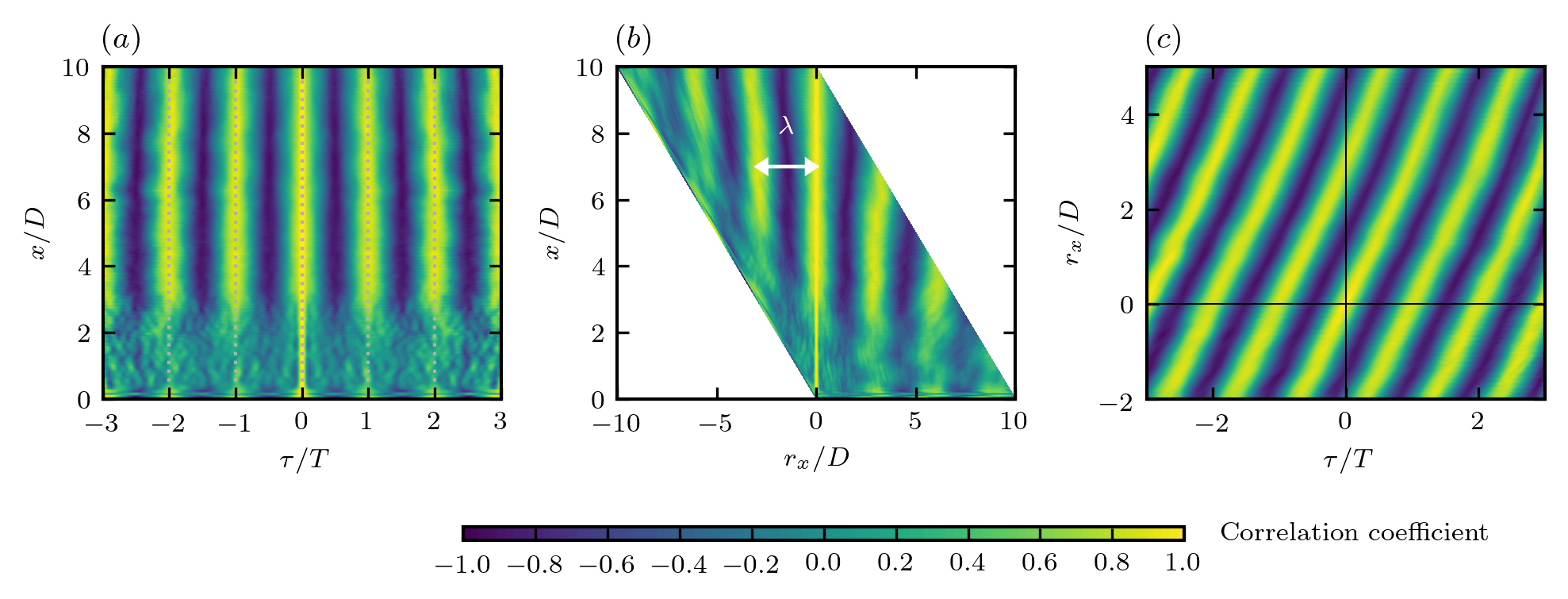}
    \caption{The space and time correlation of the transverse velocity in the wake behind the moving wind turbine: (a) the time correlation, (b) the space correlation, and (c) the space time correlation at point $x = 5D$.}
    \label{fig:spaceTimeCorrelation}
    
\end{figure}

To further understand the spatial and temporal evolution of the meandering induced by the side-to-side motion, the space and time correlations of the transverse velocity fluctuations in the wake of the oscillating wind turbine are examined in figure \ref{fig:spaceTimeCorrelation}. As seen, they are distinctively different from the space-time correlation in fully developed turbulent flows  \citep{he2017space,wu2017characteristics,wu2020,wu2021}, with clear periodicity in both space and time showing the dominant effects of the coherent wake meandering induced by the side-to-side motion. The periodicity observed on the time correlation contour at different streamwise locations as shown in figure \ref{fig:spaceTimeCorrelation} (a) agrees well with the imposed motion. The space correlations at different downstream locations are shown in figure \ref{fig:spaceTimeCorrelation} (b). Interestingly, it is observed that the wavelength ($\lambda$)  obtained from the space correlation gradually grows while traveling downstream, which is probably caused by the recovery of the streamwise velocity in the far wake, which transports the flow structures at a higher velocity. The space-time correlation is shown in figure \ref{fig:spaceTimeCorrelation} (c). Contours of straight lines with a nearly constant slope are observed, indicating the validity of Taylor's frozen flow hypothesis for the simulated cases. By inspecting figure \ref{fig:spaceTimeCorrelation} (c) carefully, it is seen that the slope becomes slightly steeper when traveling downstream, which indicates the increase of the streamwise convection velocity, being consistent with the increase of wavelength shown in figure \ref{fig:spaceTimeCorrelation} (b). 
\subsubsection{Time-averaged wake characteristics}
\begin{figure}
    \centering
    \includegraphics[width=\textwidth]{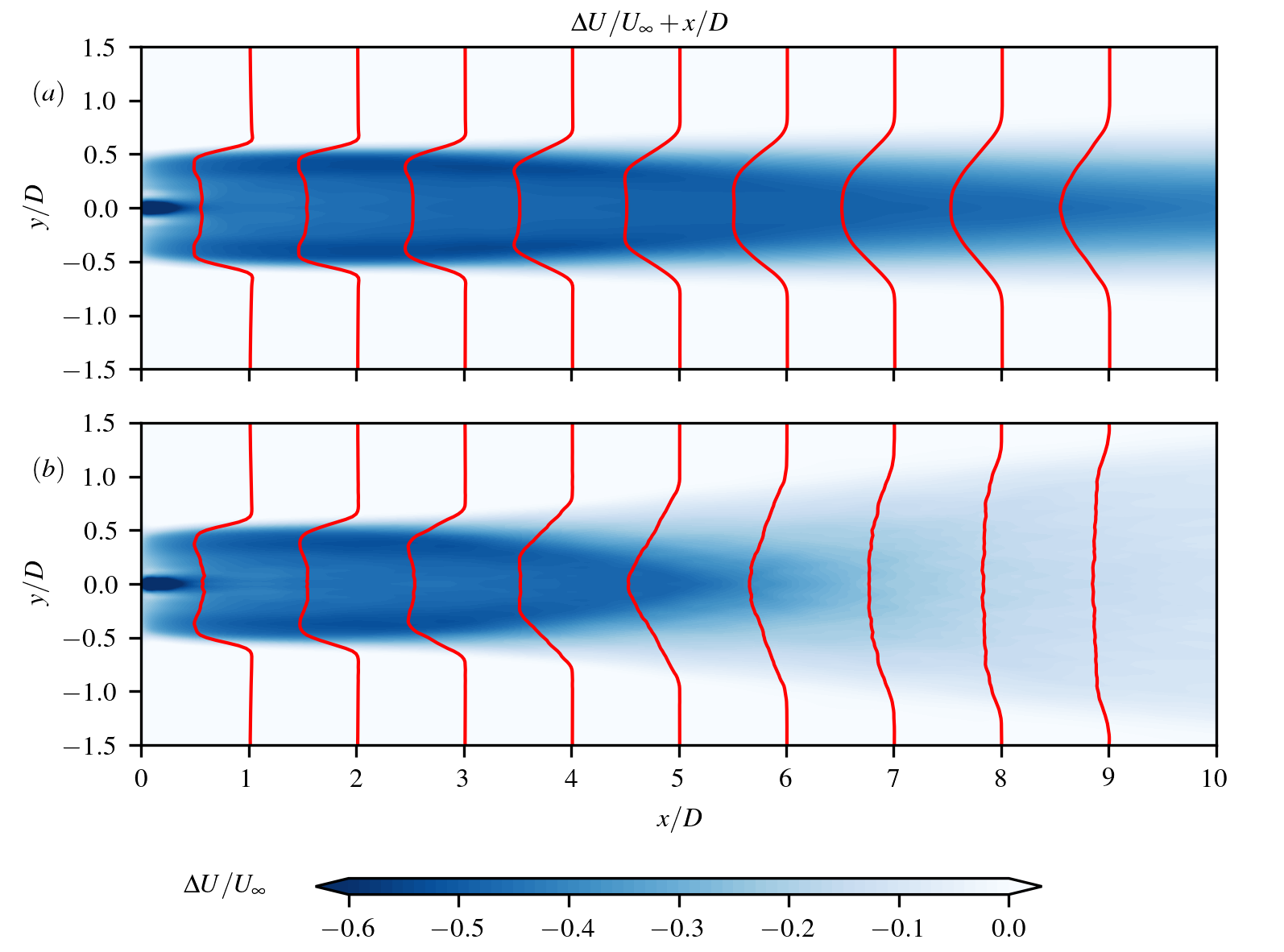}
    \caption{Time-averaged streamwise velocity deficit contour in the wake behind a fixed wind turbine (a) and a wind turbine subject to side-to-side simple harmonic motion with $\textrm{St} = 0.25$ and amplitude $A/D = 0.04$ (b). The red lines represents the spanwise variation of the velocity deficit at different downstream locations.  \label{fig:timeAveragedWakeCompare} }
\end{figure}
\begin{figure}
    \centering
    \includegraphics[width=\textwidth]{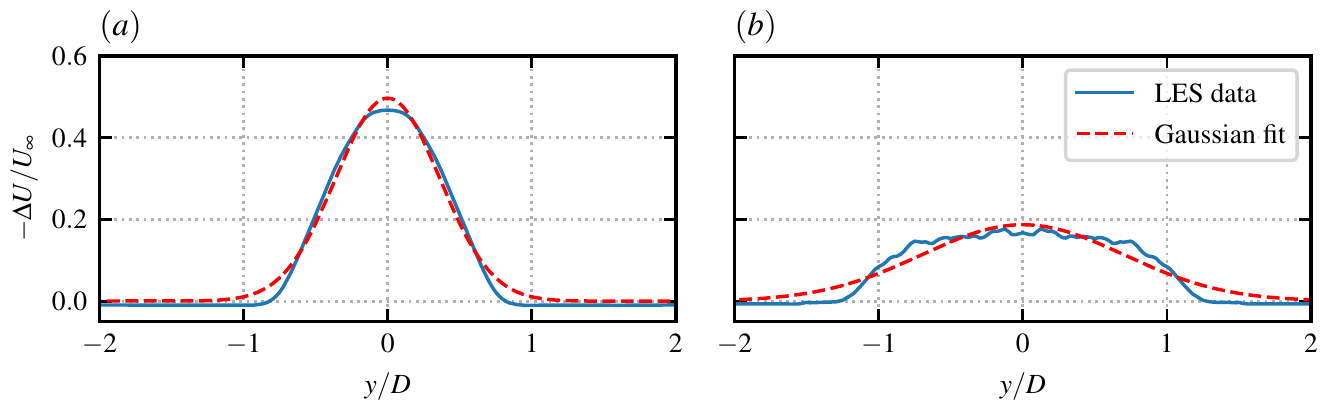}
    \caption{Spanwise profile of the velocity deficit at $x=8D$ behind the fixed wind turbine (a) and the oscillating wind turbine (b).}
    \label{fig:velocityDeficitCompareWithGaussianFit}
\end{figure}

The effects of turbine motion on the time-averaged wake are examined in this section. First, we show the contour of streamwise velocity deficit in figure \ref{fig:timeAveragedWakeCompare}. {\color{black} It is observed that the time-averaged velocity deficit of the fixed and oscillating cases are close to each other in the near wake region ($x<3D$). At further downstream locations, the wake of the fixed wind turbine as shown in figure \ref{fig:timeAveragedWakeCompare} (a) only shows a mild wake expansion and the wake shear layer gradually grows in both inward and outward directions. The maximum deficit along the wake centerline remains approximately $\Delta U = U- U_\infty = -0.5U_\infty$ at $x=8D$~ (a common spacing between two turbines in a wind farm  \citep{hansen2012HornsRev,yang2018large,ahsbahs2020wind}). By contrast, the wake behind the oscillating wind turbine as shown in figure \ref{fig:timeAveragedWakeCompare} (b) shows an obvious broader wake expansion starting from $x=4D$, as a result of large-scale meandering motion which spreads the low speed region in an larger extend. A direct result of this broad wake expansion is the faster recovery of momentum deficit in the wake core. As shown in figure \ref{fig:velocityDeficitCompareWithGaussianFit}, the time-averaged velocity deficit is approximately $\Delta  U \approx -0.2U_\infty$ at $x=8D$, being significantly lower than that behind the fixed wind turbine  ($\Delta U \approx -0.4U_\infty$). From the practical point of view, this faster wake recovery is beneficial for turbines in the wake because of the increase of the available wind power. It is also worth noting that the spanwise profile of the velocity deficit in the far wake is different from the Gaussian bell shape, especially for the oscillating wind turbine, whose wake profile is featured by a relatively flat central part and a steep transition near the wake boundary and related with the PDF of the instantaneous wake center positions as shown in the previous subsection. } 

\subsubsection{Turbine-added turbulence}

\begin{figure}
    \centering
    \includegraphics[width=\textwidth]{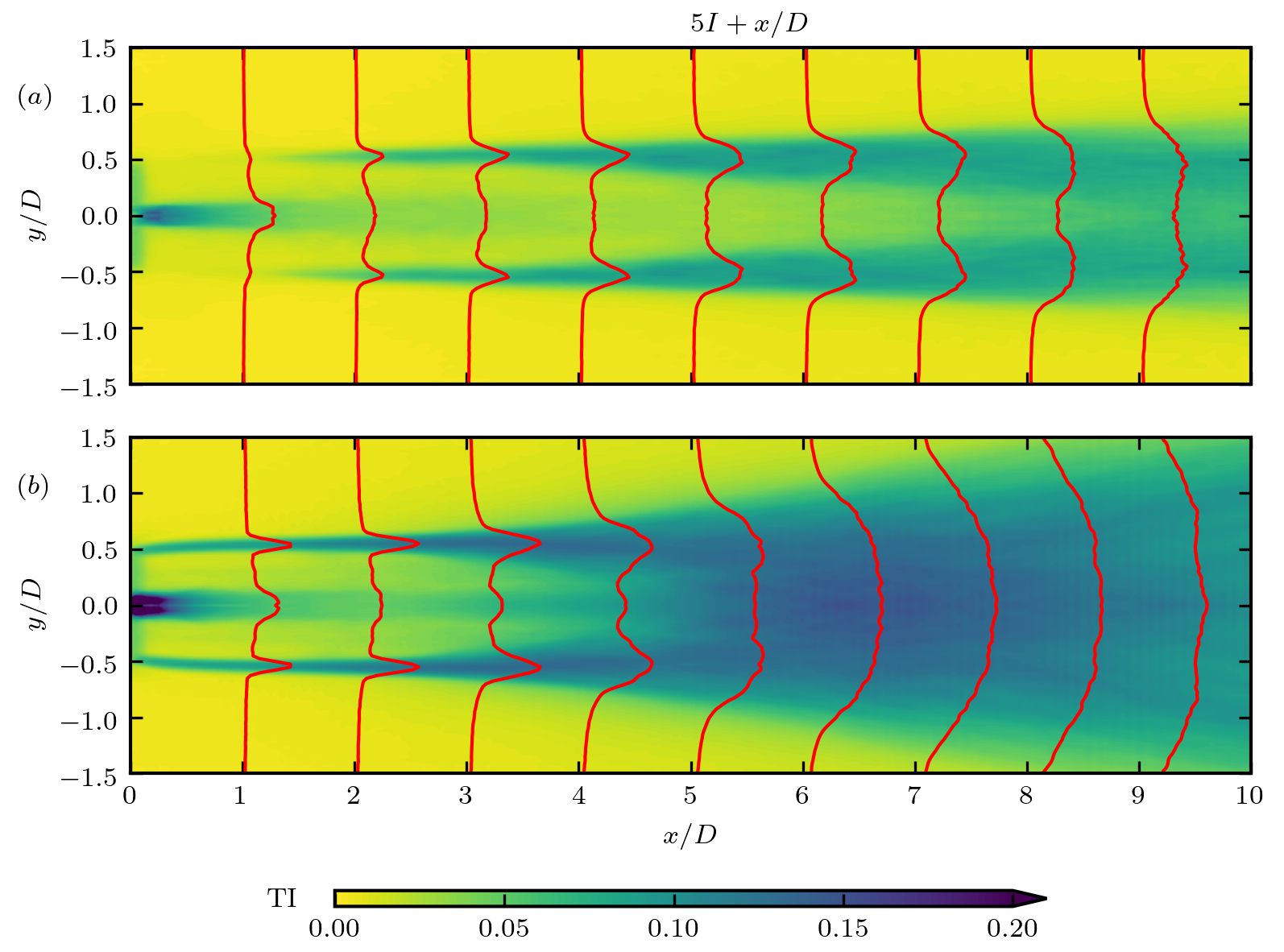}
    \caption{Turbulence intensity (TI) in the wake behind a fixed wind turbine (a) and a wind turbine subject to side-to-side simple harmonic motion with $\textrm{St} = 0.25$ and amplitude $A/D = 0.04$ (b). The transverse profiles of TI are plotted with red lines at different downstream locations.}
    \label{fig:turbulenceIntensityCompare}

\end{figure}

Except for the benefit of the faster wake recovery,  the wake meandering also results in stronger unsteady loads for the downstream turbines, which is unfavorable for the structure of the downstream turbines. To assess this effect, the turbulence intensity (TI) in the wake of the oscillating wind turbine, which is defined  as $\displaystyle \textrm{TI} = \sqrt{\frac{1}{3}(u^{\prime2} + v^{\prime2} + w^{\prime2})}/U_\infty$, where $u^{\prime},~v^{\prime},w^{\prime}$ are the root-mean-square of the fluctuation of the three velocity components, is computed and compared with the fixed turbine case. The obtained contours and profiles of the TI are shown in figure \ref{fig:turbulenceIntensityCompare}. As expected, the oscillating motion of wind turbine has drastically intensified the turbulence intensity in the wake in terms of both magnitude and spatial extent. {\color{black} Specifically, it is observed that the TI behind the hub and the tip of the rotor is intensified in the near wake because of turbine oscillation. Starting from $x=2D$, the TI grows in the shear layer of the wake, showing a bi-modal distribution in the spanwise direction, and extends in both inward to the wake core and outward to the free stream as traveling downstream. At approximately $x=5D$, the shear layers from the tips meet at the wake center for the oscillating turbine case, which happens at nearly $x=9D$ for the fixed turbine case.  At further turbine downstream locations, the TI ceases to grow and slightly decrease starting from $x=7D$. At $x=8D$, the TI in the oscillating turbine case is also significantly higher when compared with the fixed turbine case, with the maximum TI $\approx$ 13.4\%, 8.5\% for the former and latter cases, respectively. It is noticed that this level of turbulence intensity is still within the range of the wind turbine design standard (turbulence category C as per IEC 61400-1) and comparable to the atmospheric turbulence intensity caused by complex terrain topology  \citep{yang_3Dhill_2015, yang2018large, li2021wall} for land-based wind turbines \citep{DNV_Standard}}.    
\subsection{Effects of motion frequencies and amplitudes\label{sect:result2}}

After showing the results from a specific case, for which the wake meandering is induced by a given side-to-side motion, in this section we perform local LSA and carry out simulations with a series of frequency and amplitude combinations to examine how the onset of wake meandering depends on the frequencies and amplitudes of the side-to-side motions. 

\subsubsection{Local linear stability analysis}

Previous works on wind turbine wakes in the literature have demonstrated its convective instability  \citep{iungo2013linear,sarmast2014mutual,mao2018far,gupta2019low}, \ie{}, the wake behaves as an amplifier of upstream disturbances. For a prescribed base flow, the linear stability theory can predict the spatial amplification rate for an infinitely small disturbance at different excitation frequencies. In the following, the local LSA is performed to assess the response of wake to small disturbances with different frequencies. The time-averaged flow behind a fixed wind turbine is simplified using locally axis-parallel and axisymmetric assumptions, \ie{}, the velocity is simplified as $(\overline U(r),~0,~0)$ in the streamwise, radial and azimuthal directions (where $\overline{U}$ is the temporally and azimuthally averaged streamwise velocity from the fixed turbine case), and employed as the base flow for the LSA. The perturbation velocity components $(u_x,~u_r,~u_\theta)$  and the perturbation pressure $(p)$ are assumed to be the form of $(u_x,~u_r,~u_\theta,p) = (\hat{u}_x,~\hat{u}_r,~\hat{u}_\theta,~\hat{p}) \left(\exp( ik x  + in\theta - i \omega t)\right )$, where $\hat{\cdot}$ dentoes the shape function in the radial direction. Spatial stability analyses are performed with the real angular frequency $ \omega = 2\pi f $ and the complex wave number $k=k_r+ik_i$. Substituting the perturbation variables into the linearized Navier-Stokes equations yields,  
\begin{align}
   i k \hat{u}_x + \frac{\hat{u}_r}{r} + \frac{d \hat{u}_r}{d r} + \frac{ i n  \omega}{r} \hat{u}_\theta & = 0 \\
   i ( k \overline{U} -\omega ) \hat{u}_x + \frac{d \overline{U}}{d r} \hat{u}_r  & = - i k \hat{p}, \\
   i ( k \overline{U} - \omega ) \hat{u}_r & = - \frac{d \hat{p}}{d r}, \\
   i ( k \overline{U} - \omega ) \hat{u}_\theta & = - \frac{i n}{r} \hat{p}. 
\end{align}
The above equations are the same as those derived by Batchelor \& Gill  \citep{batchelor1962analysis} for axisymmetric jets/wakes, expect that the viscous terms are excluded, because their contribution is less important for base flow velocity profiles with inflection points  \citep{schmid2002stability}. The azimuthal wave number is selected as $n=1$, because a side-to-side motion velocity ($u_y$) can be projected in the radial and azimuthal directions as $u_r = u_y \cos\theta$ and $u_\theta = u_y \sin\theta $.  The linearized Navier-Stokes equations are discretized with Chebyshev polynomials on $128$ Gauss–Lobatto nodes $\xi \in \left[-1,1 \right]$ and stretched  to $r \in [0, r_\textrm{max}]$ through an algebraic mapping  \citep{schmid2002stability} defined as $\displaystyle r = a \frac{1+\xi}{b-\xi}$ with $\displaystyle a = \frac{r_\textrm{max}}{r_\textrm{max}-2}$ and $\displaystyle b=1+\frac{2a}{r_\textrm{max}}$. The size of the computational domain is chosen as $r_\textrm{max} = 15D$. The size of the domain and the number of discretization points are verified to be large enough to not influence the stability analyses.  
The far field boundary condition is that the perturbations vanish at $r = r_\textrm{max}$. On the wake centerline, $\hat{u}_x = 0$ $\hat{p}=0$ and $\hat{u}_r + i \hat{u}_{\theta} = 0$ are imposed following Batchelor \& Gill  \citep{batchelor1962analysis}. 
\begin{figure}
    \centering
    \includegraphics[width=\textwidth]{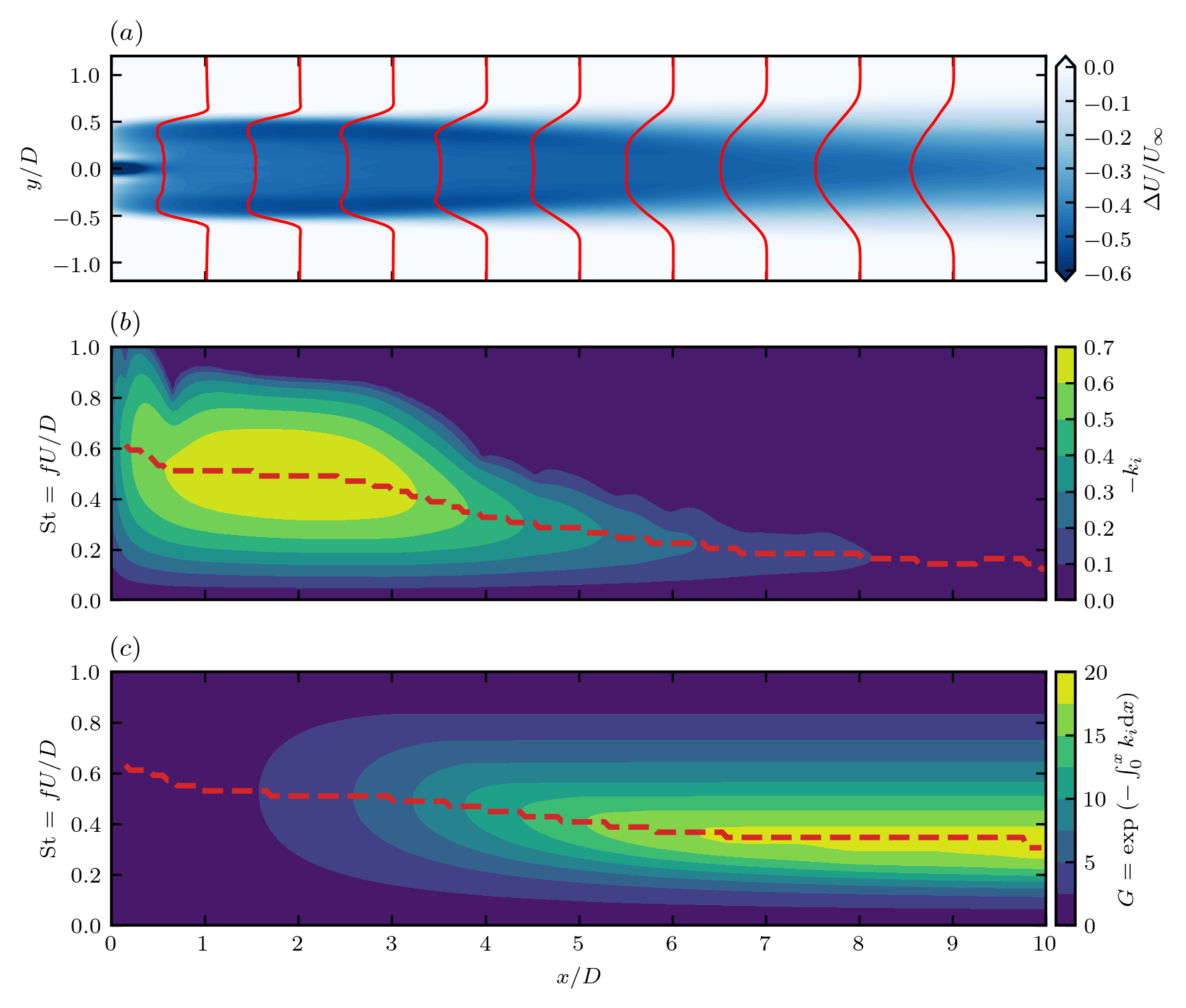}
    \caption{Local linear stability analysis results of the wake behind the fixed wind turbine: $(a)$ the streamwise velocity deficit, $(b)$ contours of local spatial growth rate $-k_i$, and $(c)$ the integrated amplification factor  $\displaystyle G = \exp{(- \int_0^x k_i \textrm{d}x)}$. The red dotted lines in $(b)$ and $(c)$ indicate the most amplified frequency along the streamwise locations. }
    \label{fig:linearStability}

\end{figure}

Figure \ref{fig:linearStability} shows the results of local LSA conducted at different downstream locations. As seen in figure \ref{fig:linearStability} (a), the velocity deficit profiles in the region between the core of the wake and the free stream is characterized by a shear layer. The thickness of this shear layer gradually increases in the turbine downstream direction due to the momentum mixing between the wake and the freestream, and consequently makes the local stability properties vary with the downstream distance as shown in figure \ref{fig:linearStability} (b). In the near wake  ($x<3D$), the steep velocity variation in the shear layer results in a broad range of unstable frequencies $\textrm{St} \in (0, 0.9)$ approximately, with the most unstable frequency at $\textrm{St} \approx 0.5$. At further turbine downstream locations, both the most unstable frequency and the range of the unstable frequencies decrease. This trend of decrease of the most unstable frequency via the downstream distance is in agreement with the work of Gupta \& Wan  \citep{gupta2019low}, where LSA is applied to the time-averaged wake of a tidal turbine. Such downstream variations of the spatial growth rate indicates that the most unstable perturbations in the near wake becomes stable when propagating downstream and ceases to grow subsequently. On the other hand, perturbations with lower frequencies may excel in the far wake since they can grow for a longer distance. To demonstrate this accumulation effect via downstream distances, figure \ref{fig:linearStability} (c)  displays the contours of the integrated amplification factor defined as $ G =\displaystyle \exp{(-\int_0^x k_i \textrm{d} x )}$. It is found that the frequency with the largest amplification factor also decreases in the streamwise direction, being consistent with that of the local growth rate $-k_i$. In the far wake, perturbations with evident amplifications fall in the frequency range of $0.1<\textrm{St}<0.8$. The strongest amplification factors are predicted for $0.2<\textrm{St}<0.4$, explaining the large-scale wake meandering triggered by turbine side-to-side motion with $\textrm{St}=0.25$ as presented in section \ref{sect:result1}. Furthermore, this frequency range is rather low compared to the commonly encountered excitement frequencies generated by wind turbines, \eg{}, the rated rotor frequency and the blade passing frequency are approximately equal to $0.2 \textrm{~Hz}$ ($\textrm{St} \approx 2.2 $) and $0.6 \textrm{~Hz}$ ($\textrm{St} \approx 6.6 $), respectively for the NREL 5WM wind turbine  \citep{jonkman2009definition} as well as other wind turbines at utility scale. As a result, unstable modes of the wake are unlikely to be triggered by the rotation of blades alone, being consistent with the literature that the hub vortex at lower frequencies plays an important role in triggering the shear layer instability and the far wake meandering for the bottom-fixed turbine  \citep{kang2014onset}. Moreover, this range is also lower than the wave frequencies at operational sea states ($\textrm{St} \ge 1.0$) as shown in table \ref{tab:frequencies}, indicating that floating wind turbines moving at wave frequencies cannot trigger significant wake meandering which has been shown by experiments conducted by Schliffke \etal{}  \citep{schliffke2020wind}. 
On the other hand, this low frequency range ($0.2<St<0.4$) overlaps partially with the natural frequencies of the roll or the sway motions as shown in table \ref{tab:frequencies}. 

Although the above LSA can provide useful information on the instability properties of the wake, it is worth noting the analyses are carried out based on several assumptions for the base flow and disturbances, \ie{}, i) the axis-parallel and axisymmetric assumption for the base flow, ii) the small perturbation assumption for linearizing the  Navier-Stokes equations, and iii) the inviscid flow assumption. Often, these prerequisites are not fully satisfied in real applications, for which the base flow is complex and the disturbances are of finite amplitudes, such that the nonlinear effects may manifest. To further investigate the wake characteristics of the oscillation wind turbine with side-to-side motions of finite amplitude and to verify the results from the LSA, results from LES cases with the frequencies and amplitudes with setups presented in tables \ref{tab:amplitudes} and \ref{tab:frequencies} will be analyzed in the following subsections.

\subsubsection{Instantaneous wake}
\begin{figure}
    \centering
    \includegraphics[width=\textwidth]{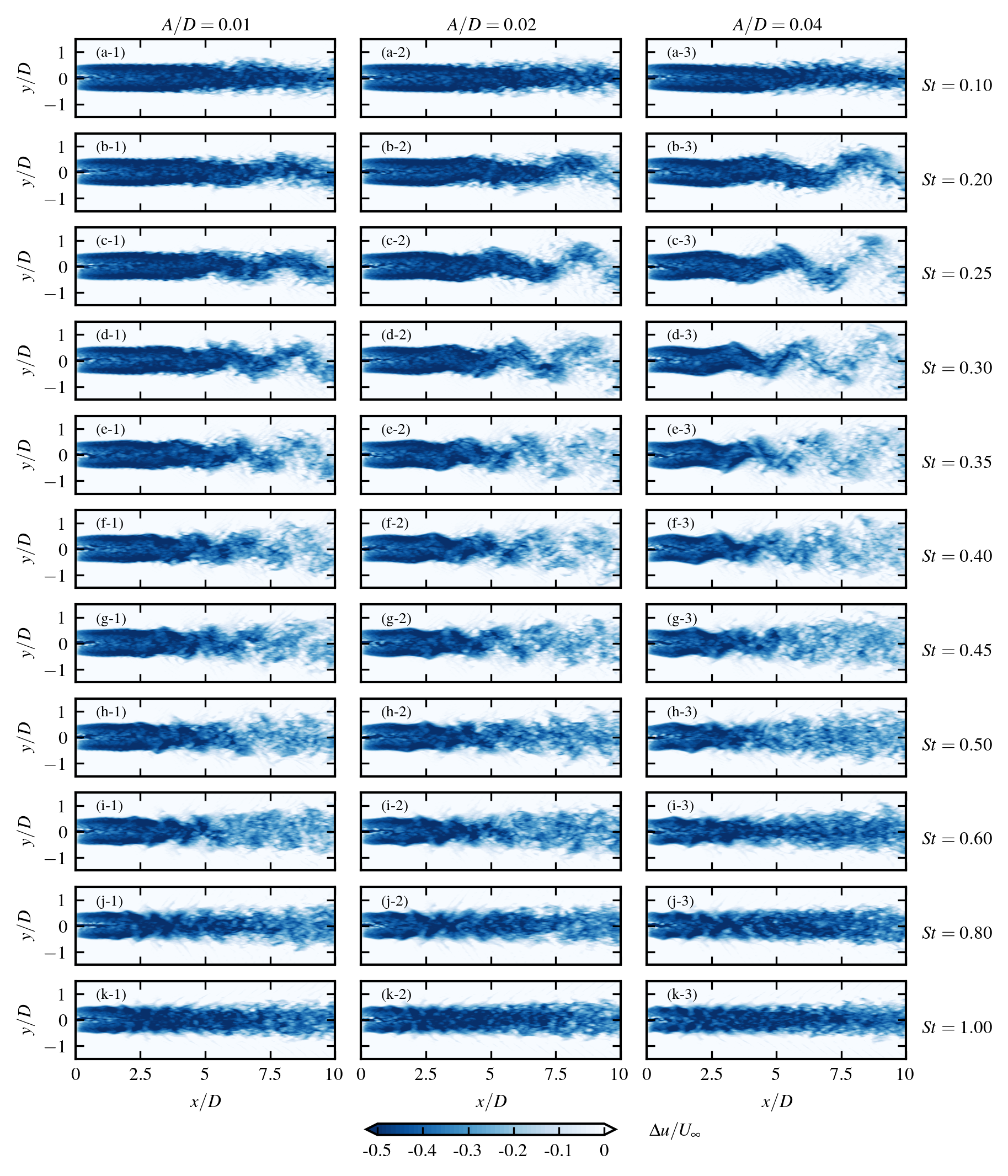}
    \caption{The instantaneous streamwise velocity deficit behind a wind turbine subject to side-to-side simple harmonic motion for various frequencies and amplitudes. Figures in the same row have the same motion frequency denoted by the Strouhal number given in the last column. Figures in the same column have the same motion amplitudes. }
    \label{fig:wakeMeanderingVelocityFields}
\end{figure}
Snapshots of the instantaneous streamwise velocity deficits are plotted for comparison in figure \ref{fig:wakeMeanderingVelocityFields}. As observed, the wake behavior varies via the turbine oscillation frequency, being consistent with the LSA. For cases with the lowest and the highest motion frequencies, \ie{}, for $\textrm{St} = 0.1$ and $\textrm{St} \ge 0.8$, no apparent wake meandering is observed and the wake is almost straight. For the cases with the lowest motion frequency ($\textrm{St} = 0.1$), the spanwise motion of the wake is very mild and the wake is close to that behind a fixed turbine. Although increasing the amplitude of the side-to-side motion to $A/D=0.04$ can deform slightly the wake into a sinusoidal form, the amplitude of wake center displacement is still much smaller than the wake width. For cases with the largest frequencies ($\textrm{St} \ge 0.8$), the wake centerline is approximately straight and without significant oscillations, indicating the overall meandering motion of the wake is negligible. On the other hand, small-scale fluctuations are observed on the wake boundary, which, however, do not show apparent amplifications as travelling downstream and remain locally near the boundary without apparent interaction with the core of the wake.  Moreover, it is observed that these small scale fluctuations do not seem to grow in magnitude when increasing the amplitude of the side-to-side motion that roughly the same instantaneous wakes are observed for cases at $\textrm{St} \ge 0.8 $ with different amplitudes considered in the present study. In contrast, the cases with $0.20 \le \textrm{St} \le 0.60 $ show a much stronger far wake meandering, which is consistent with the larger amplification factor predicted by LSA. Specifically, the cases with $0.20 \le \textrm{St} \le 0.40$ exhibit apparent sinusoidal form deformations with a continuous growth of wake meandering amplitudes via the downstream distance until the very far wake. Larger meandering amplitudes are also observed when the motion amplitudes is larger with the same turbine motion frequency. For cases with higher Strouhal numbers ($0.4 < \textrm{St} \le 0.6$), however, the part of the wake with sinusoidal deformation is limited to a certain downstream distance, beyond which the wake loses this periodicity and the coherent structures cannot be identified visually,  
and the effects of motion amplitude on the wake meandering in the far wake are hard to tell directly from these instantaneous snapshots in that region because of the complex far wake flow.  

\begin{figure}
    \centering
    \includegraphics[width=\textwidth]{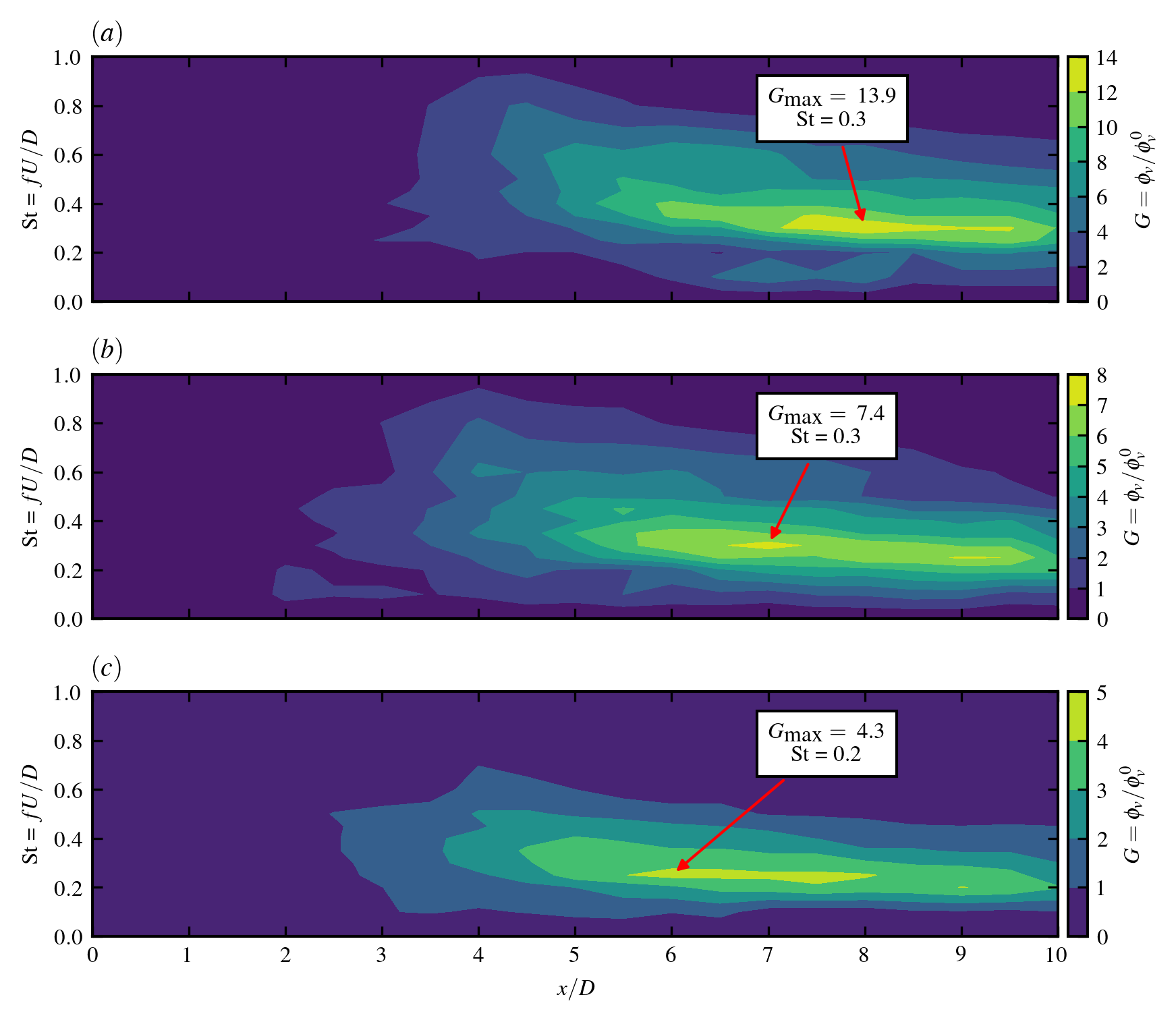}
    \caption{{\color{black} Contours of the amplification factor $G$ computed from the LES results for different turbine motion amplitudes: $(a)$ $A/D =0.01$, $(b)$ $ A/D=0.02$, $(c)$ $A/D = 0.04$. The amplification factor is defined as $G = \phi_v/\phi_v^0$, where $\phi_v$ is the amplitude of spanwise velocity in the wake obtained with discrete Fourier transform of the time series of the spanwise velocity obtained from LES along the time-averaged wake centerline, and  $\phi_v^0$ is the velocity amplitude of the wind turbine's simple harmonic motion. The largest amplification factor is marked together with its Strouhal number. }}
    \label{fig:amplificationFactor_LES}
\end{figure}

{
\color{black}
It is observed from the instantaneous results that the wake meandering is affected by both amplitude and frequency of the wind turbine oscillation. To further investigate such influences quantitatively, we compute the amplification factor of the spanwise velocity in the wake for different turbine oscillation frequencies and amplitudes, as shown in figure \ref{fig:amplificationFactor_LES}.  The amplification factor is defined as $G = \phi_v/\phi_v^0$, with $\phi_v$ the amplitude of spanwise velocity at the turbine motion frequency obtained with discrete Fourier transform from simulation results collected along the time-averaged wake centerline and $\phi_v^0$ is the velocity amplitude of the wind turbine's simple harmonic motion computed as $\phi_v^0 = 2\pi f A$, with $A$ the turbine motion amplitude and $f = U_\infty \times  \textrm{St}  / D $ the turbine oscillation frequency. The amplification factors at every $0.5D$ for $x\in [0D, 10D]$ are computed.

The obtained amplification factors are presented in figure \ref{fig:amplificationFactor_LES} $(a), (b)$, and $(c)$ for the cases with turbine motion amplitudes of $A/D=0.01, 0.02, 0.04$, respectively. As seen, all the three figures show a similar pattern for the regions with amplification factors higher than one, indicating that the initial spanwise velocity is amplified while traveling downstream for those cases. It is observed that the amplification of the spanwise velocity is most intensive for turbine motion with frequency in the range of $\textrm{St} \in [0.2,0.3]$ and gradually weakened for lower or higher frequencies. For the cases with the smallest turbine motion amplitude, \ie{}, $ A = 0.01D$ in figure \ref{fig:amplificationFactor_LES}, the location and size of the region with high amplification factors are closer to that predicted by the local LSA (as shown in figure \ref{fig:linearStability} $(c)$) when compared with the cases with higher oscillation amplitudes. In terms of the magnitudes of the amplification factors, the LES predictions are observed being in general lower than the LSA predictions, because that the hypotheses employed in LSA, \ie{}, the infinite small perturbation, axis-parallel base flow and the inviscid fluid, are not full satisfied. Moreover, it is observed that the differences between LSA and LES predictions are enlarged when increasing turbine motion amplitudes as shown in figure \ref{fig:amplificationFactor_LES} $(b)$ and $(c)$, which suggests that the non-linearity caused by the finite size perturbations plays a role on reducing the growth rate, being consistent with the nonlinear stability theory  \citep{schmid2002stability} and a previous work on tidal turbine wakes  \citep{gupta2019low}. Furthermore, figure \ref{fig:amplificationFactor_LES} shows that the increase of turbine motion amplitudes slightly lowers the most amplified frequency and narrows the region with high amplification factors. 
}

\subsubsection{Time-averaged wake}  

\begin{figure}
    \centering
    \includegraphics[width=\textwidth]{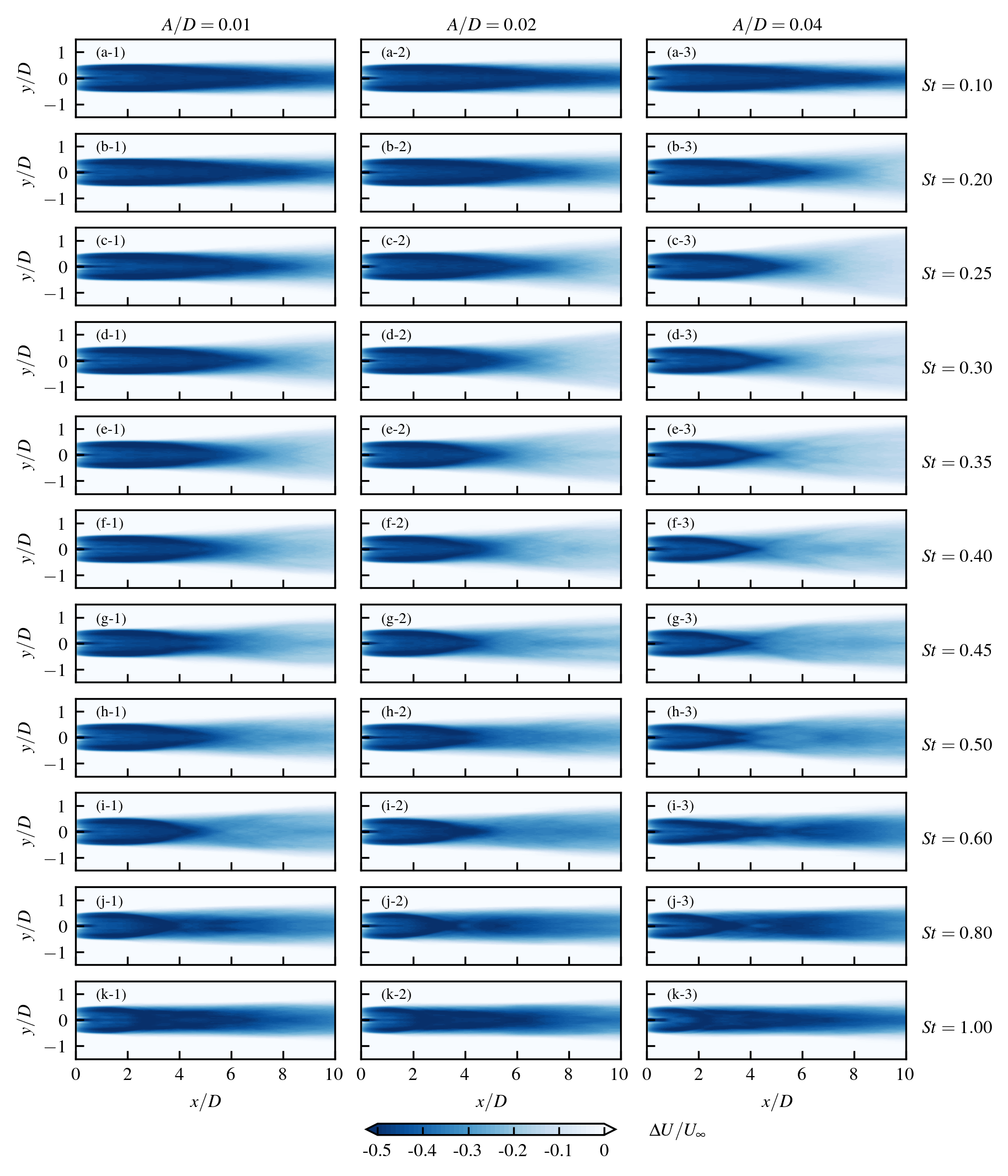}
    \caption{The time-averaged streamwise velocity deficit at the hub-height plane. The wind turbine is subject to side-to-side simple harmonic motion defined with various frequencies and amplitudes. Figures in the same row have the same motion frequency denoted by the Strouhal number given in the last column. Figures in the same column have the same motion amplitudes.}
    \label{fig:timeAveragedWakeAll}
\end{figure}

\begin{figure}
    \centering
    \includegraphics[width=\textwidth]{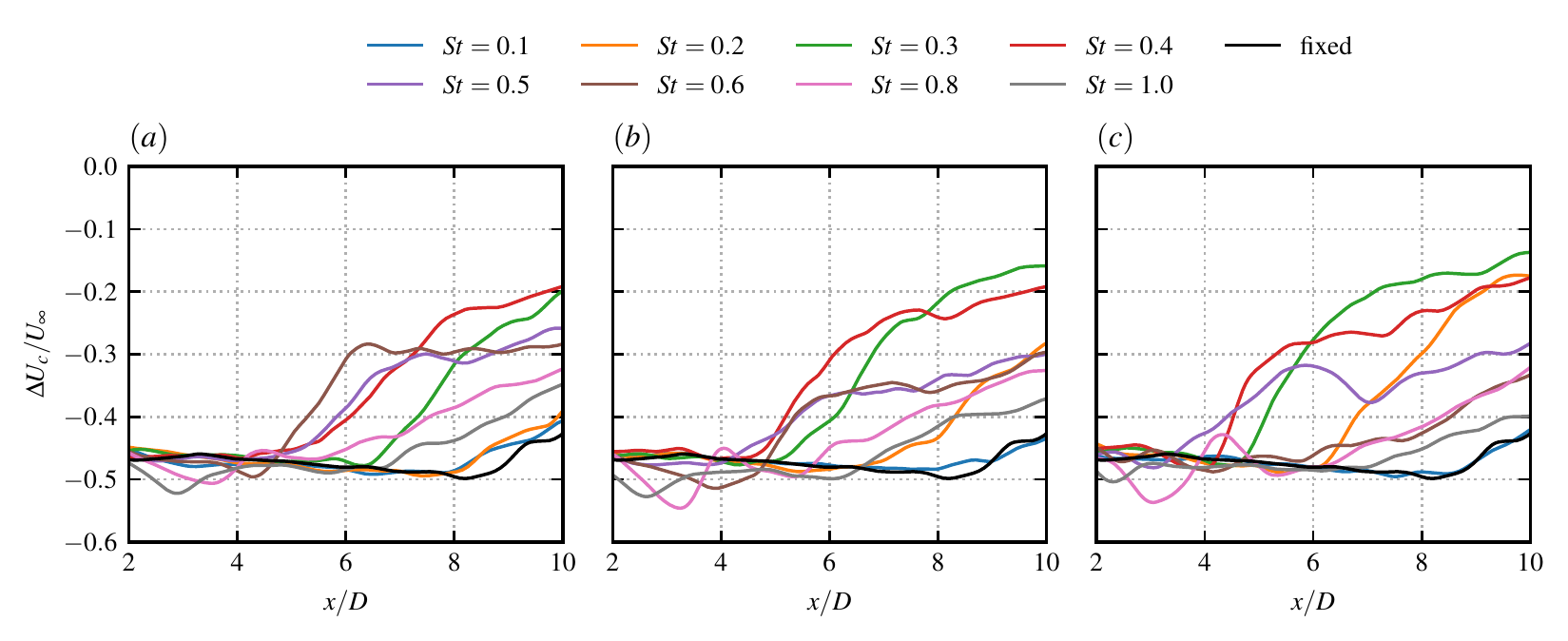}
    \caption{The time-averaged centerline velocity deficit for cases with different turbine oscillation frequencies and motion amplitudes of $A/D = $ 0.01, 0.02, and 0.04 for subplots (a),(b),and (c), respectively. }
    \label{fig:timeAveragedCenterlineVelocity}
\end{figure}

To investigate the influence of turbine motion on the wake recovery, we analyze the time-averaged velocity deficits computed from different cases. In figure \ref{fig:timeAveragedWakeAll}, we show contours of the time-averaged streamwise velocity deficits. As seen, the recovery of the time-averaged wake for cases with $\textrm{St} = 0.1$ and $\textrm{St} \ge 0.8$ is slow and the wake is close to that of a fixed wind turbine, being consistent with the instantaneous velocity deficit. In contrast, the velocity deficit contours are significantly altered for cases with motion frequencies in the range of $0.20 \le \textrm{St} \le 0.60$, and the wake recovery is observed being dependent on both motion amplitude and frequency. Taking the cases with $A/D=0.01$ as example, the velocity starts to recover from the wake edge before finally reaching the wake centerline. The velocity recovery on the wake boundaries is found being faster with higher platform motion frequencies, which is related to the stronger mixing induced by the steeper wavy deformation of the shear layer as seen from the instantaneous wake snapshots (figure \ref{fig:wakeMeanderingVelocityFields}). However, faster overall far-wake recovery is only observed for the cases in the frequency range of $0.2< \textrm{St} <0.6$  where intense wake meandering is observed in the far wake, indicating the large-scale wake meandering is more efficient than the mixing induced by waves in the shear-layer. 

The above observation is further confirmed by the quantitative comparison shown in figure \ref{fig:timeAveragedCenterlineVelocity} (a), where the velocity deficit along the mean wake centerline is displayed for cases with different turbine motion frequencies at the same motion amplitude $A/D=0.01$. As seen, the centerline velocity for cases with $\textrm{St}=0.6$ recovers rapidly in $4D<x<6D$  then stops growing. On the other hand, cases with lower turbine motion frequencies ($\textrm{St} = 0.3$ and $\textrm{St} = 0.4$ ) are featured by the smallest velocity deficit of $\Delta U_c/U_\infty \approx 0.2$  at $x=10D$, compared with  $\Delta U_c/U_\infty \approx 0.4$ from the fixed wind turbine case. Another interesting phenomenon found in figure \ref{fig:timeAveragedCenterlineVelocity} is the influence of turbine motion amplitude on the wake recovery is nonlinear, \eg{}, the turbine motion frequency with strongest wake recovery in the far wake is different when the motion amplitude increase from $A/D = 0.01$ to $A/D = 0.04$. 



\subsubsection{Turbine-added turbulence}

\begin{figure}
    \centering
    \includegraphics[width=\textwidth]{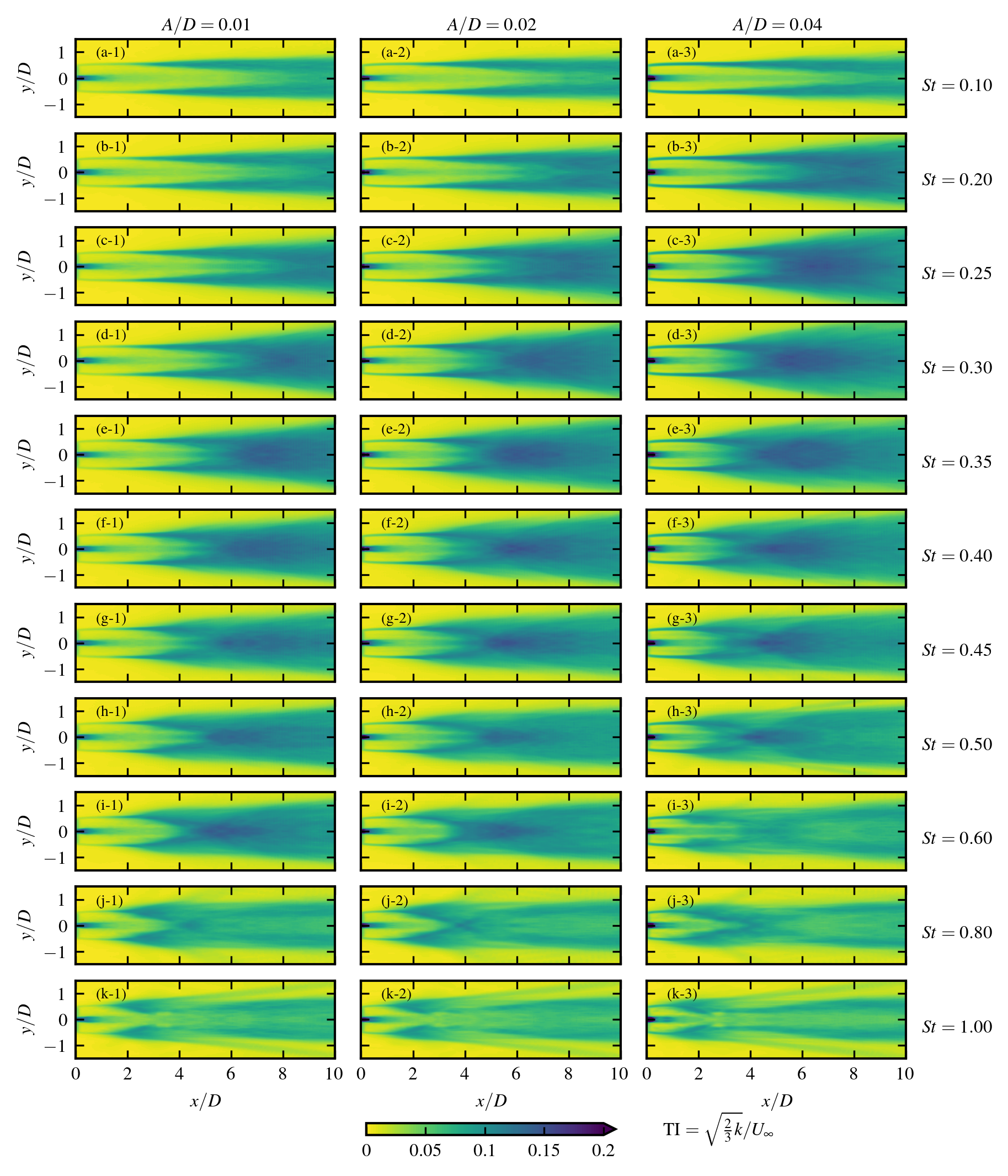}
    \caption{The turbulence intensity at the hub-height plane. The wind turbine is subject to side-to-side simple harmonic motion defined with various frequencies and amplitudes. Figures in the same row have the same motion frequency denoted by the Strouhal number given in the last column. Figures in the same column have the same motion amplitudes.}
    \label{fig:wakeTKEAll}
\end{figure}

\begin{figure}
    \centering
    \includegraphics[width=\textwidth]{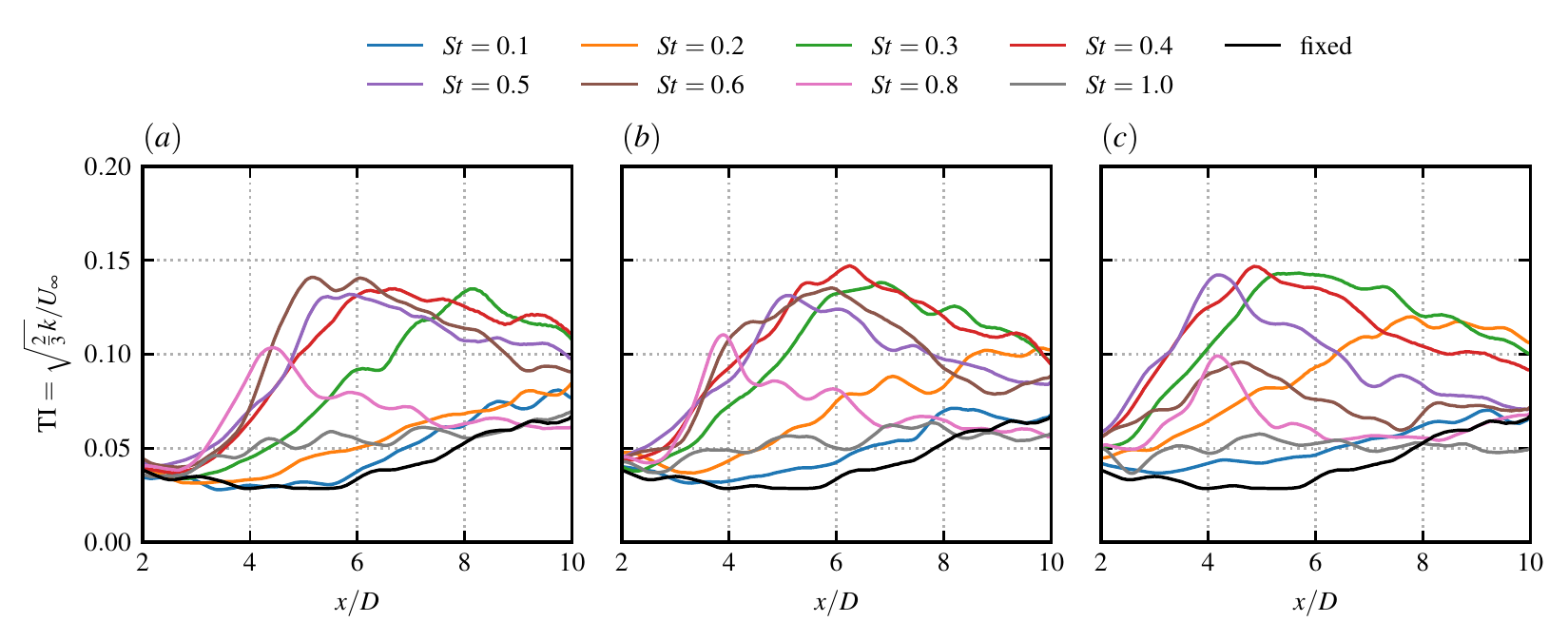}
    \caption{The turbulence intensity on the wake centerline for cases with different turbine oscillation frequencies and motion amplitudes of $A/D = $ 0.01, 0.02, and 0.04 for subplots (a),(b),and (c), respectively. }
    \label{fig:centerlineTKE}
\end{figure}

In this section, we examine the characteristics of turbine-added turbulence for cases with different side-to-side motions. Figure \ref{fig:wakeTKEAll} displays the contours of the turbulence intensity (TI) on the hub-height plane for cases with different motion amplitudes and frequencies. For the cases with $\textrm{St}=0.10$, both magnitude and spatial distribution of TI are very similar to that of the fixed turbine case as shown in figure \ref{fig:turbulenceIntensityCompare} (a), \ie{}, the TI appears first on the wake edge, making a bi-modal spatial distribution then it gradually expands inwards to the wake centerline and outwards to the freestream. For higher turbine motion frequencies $0.2  \le \textrm{St} \le 0.6$, the TI is generally higher than the fixed wind turbine and spreads in a larger extent. For $\textrm{St} \ge 0.8$, the TI becomes weaker due to the absence of large-scale wake meandering as observed in the instantaneous wake snapshots. Figure \ref{fig:centerlineTKE} compares the TI along the wake centerline for cases with different turbine motion amplitudes and frequencies. In figure \ref{fig:centerlineTKE} (a), it is found that the streamwise variations of TI for cases with $\textrm{St}=0.1$, $\textrm{St}=0.2$ and $\textrm{St}=1.0$ are close to the case of the fixed wind turbine. Other cases, on the other hand,  exhibit a much stronger streamwise increase of TI before reaching their peak, after which the TI starts to decrease. The location of peak shift towards the turbine with increasing turbine oscillating frequency. For all cases, the peak TI does not exceed the upper limit of 0.15 and decreases to approximately TI$=0.1$ at $x=10D$. Increasing the turbine oscillation amplitudes does not amplify the magnitude of TI significantly, while shifts the peak of TI upstream and it leaves a longer distance for the TI to decrease before reaching possible neighbours in the downstream.  Consequently, the final TI reaching these downstream turbines located in the far wake ($x/D=10$) can be smaller for cases with larger amplitudes for cases with certain side-to-side motion frequencies, \eg{}, the cases with $0.3 \le \textrm{St} \le 0.5$. Such decrease is probably related to the recovered velocity deficit in the wake, which makes the nominal velocity fluctuations induced by the wake meandering smaller.        
\section{Discussion}

The most important finding from this work is that the turbine side-to-side motion can be a novel origin for the onset of wake meandering. This finding complements the existing wake meandering mechanism   \citep{iungo2013linear,mao2018far, gupta2019low}, and is of particular importance for FOWTs because their compliant foundations are easily subject to motions induced by waves/winds. Platform motions at certain frequencies can be amplified in the far wake due to wake instability and leads to wake meandering with amplitude comparable to the turbine rotor diameter.  {\color{black} As revealed by the space-time correlation analysis, the frequency of the meandering motion is strongly related with that of the side-to-side motion and well preserved in the entire wake, being distinctively different from the meandering motion caused by inflow large-eddies. The PDF of the instantaneous wake center positions, which is featured by two peaks, one at each extremity, being consistent with the PDF of the side-to-side FOWT motion, is significantly different from the normal distribution commonly observed for wake meandering in the literature.  
}

{\color{black} The wake meandering observed in the simulated cases with ideal setups is very likely to happen for real FOWTs.} The target wind turbine in this study (NREL 5WM reference wind turbine) is one of the most studied turbines for the FOWT research. The frequencies and amplitudes considered in the present cases are feasible in realistic applications.  The turbine motion amplitudes being less than $0.04D$, are within the design requirement for offshore wind turbines  \citep{nejad2019effect}. The frequencies resulting in the significant far wake meandering falls in the range of $0.2 \le \textrm{St} \le 0.6$, although being lower than the wave frequencies commonly encountered (corresponding to $ \textrm{St} > 1.0$ for the present turbine design), are within the range of the natural motion frequencies of the floating platform, indicating the importance of considering wake meandering induced by the FOWT motion in the design of FOWT farms. For example, the well studied reference offshore wind turbine platform, the DeepCwind semi-submersible platform  \citep{robertson_2014_offshore}, has its natural roll period about 26 s ($\textrm{St} \approx 0.4$), which will dominate the platform motion when second-order wave forces are considered  \citep{mahfouz2021response}. The present work shows that this design may trigger large-scale far-wake meandering when the platform oscillates at its natural roll frequency. On one hand, the induced wake meandering can help wake recovery. On the other hand, wake meandering at the frequency close to the platform's natural frequency may even trigger large motion of the platform of downstream turbines {\color{black}like a chain reaction which may eventually propagates in the entire wind farm.} This trade-off between the benefit and the risk must be carefully considered when designing the hydrodynamic properties of the platform.

Thirdly, the present work demonstrates that the local LSA can be employed to develop low-order models for predicting wake meandering caused by side-to-side motion of FOWTs. {\color{black} The comparison between the predictions from LSA and LES shows that the local LSA can predict the least stable frequency and the amplification factor with a reasonable accuracy, especially when the imposed turbine motion is small (\eg{}, $A/D = 0.01$). Although the amplification factors are overpredicted by LSA especially for turbine motions with high amplitudes as a result of the nonlinear effect, the local LSA employed in this work can at least give an upper bound of the meandering motion in the far wake. Besides the non-linearity effects, the inviscid fluid assumption may also lead to over predictions of the amplification factor for LSA, which can be incorporated in future studies to further improve the accuracy of the present LSA, such as including the non-linearity using the Ginzburg-Landau equation  \citep{gupta2019low}, and adding the eddy viscosity to the linearized Navier-Stokes equations as shown by Rukes \etal{} for a turbulence jet flow  \citep{rukes2016assessment}, for which the eddy viscosity could be obtained using a data-driven approach~ \citep{zhang2020regularized}.}

\section{Summary and conclusions}

The effects of the side-to-side harmonic oscillations of a floating offshore wind turbine on the wake evolution is investigated using large eddy simulations of the NREL 5MW offshore wind turbine parameterized with the actuator surface model. In the simulated cases, the side-to-side motion of the turbine with different frequencies and amplitudes are prescribed. {\color{black} The simulation results show that the side-to-side motions with frequencies in the range of $0.2 \le \textrm{St} \le 0.6 $ can trigger large-scale far wake meandering.} The wake meandering observed in the simulated cases, with its dominant frequency and the PDF of instantaneous wake positions consistent with the side-to-side motion of FOWTs, is different from the meandering observed in the literature, complementing the existing meandering mechanism that the turbine side-to-side motion can be a novel origin for the onset of wake meandering. 
Moreover, this work shows that such wake meandering induced by the FOWT side-to-side motion can be acceptably predicted using the local LSA. Using only the streamwise velocity component of the time-averaged wake behind a stationary turbine as the base flow, the local LSA is able to predict the unstable frequency range with a reasonable accuracy at a negligible computational expense. In the meanwhile, the simulation results also show that the local LSA slightly overpredicts the most unstable frequency and greatly overpredicts the amplification factors for high amplitudes of the side-to-side FOWT motion because of the nonlinear effect, which can not be considered yet in the employed local LSA. 

From the practical point of view, the side-to-side turbine motion has a two-fold influence on the downstream neighbours. On one hand, violent wake meandering in the far wake induces strong unsteady loads. On the other hand, the meandering enhances the mixing between the wake and the freestream and consequently accelerates the recovery of the time-averaged velocity deficit, which benefits the overall power output of the wind farm. Moreover, analyses using the methodologies employed in this work, \ie{}, LES and LSA, can be integrated in the design stage of FOWT. The present work shows that the wake's unstable frequencies lie over a broad range and overlaps with the floater’s natural frequencies, and thus this kind of wake meandering is highly probable with the present floater designs. Predicting the unstable frequency range of the wake as done in this work can provide an important criterion for the design of the floater's natural frequency no matter the wake meandering induced by the side-to-side turbine motion is considered as helpful or harmful.

It is worth noting that the scenario considered in the present work is rather idealized (\ie{}, only simple harmonic side-by-side turbine motion is considered and the inflow turbulence is fully ignored) to facilitate the analyses of the wake's essential features. In the real-world  applications, a FOWT is subject to complex motions of six degrees of freedom induced by both wave  \citep{liu2016investigation} and wind in the ocean atmospheric boundary layer  \citep{rockel2014experimental,wise2020wake}. These interactions between the air, the FOWT, and the waves, which can be solved using fully coupled two-phase CFD solvers  \citep{zhu2019interface,gatin2020finite,li2021spectral,zhu2021variationally}, are not considered in the present work. 

\section*{Acknowledgement}

This work is partially supported by NSFC Basic Science Center Program for ``Multiscale Problems in Nonlinear Mechanics" (NO. 11988102). 

\section*{Declaration of Interests}
The authors report no conflict of interest.

\bibliographystyle{jfm}
\bibliography{main}

\end{document}